\documentclass[twocolumn,floatfix,showpacs,prd,aps,tightenlines]{revtex4}
\usepackage{amsmath}
\usepackage{graphicx}
\usepackage{psfrag}
\usepackage{color}
\usepackage{dcolumn}
\usepackage{bm}

\newcommand{\bea}{\begin{eqnarray}}
\newcommand{\eea}{\end{eqnarray}}
\newcommand{\beq}{\begin{equation}}
\newcommand{\eeq}{\end{equation}}
\newcommand{\KMS}{\rm km\ s^{-1}}

\begin{document}

\def\fun#1#2{\lower3.6pt\vbox{\baselineskip0pt\lineskip.9pt
  \ialign{$\mathsurround=0pt#1\hfil##\hfil$\crcr#2\crcr\sim\crcr}}}
\def\lap{\mathrel{\mathpalette\fun <}}
\def\gap{\mathrel{\mathpalette\fun >}}
\def\kms{{\rm km\ s}^{-1}}
\def\vk{V_{\rm recoil}}

\title{Statistical studies of Spinning Black-Hole Binaries}

\author{
Carlos O. Lousto,
Hiroyuki Nakano,
Yosef Zlochower,
Manuela Campanelli}

\affiliation{Center for Computational Relativity and Gravitation,\\
and School of Mathematical Sciences, Rochester Institute of
Technology, 85 Lomb Memorial Drive, Rochester, New York 14623}

\begin{abstract}
We study the statistical distributions of the spins of generic
black-hole binaries during the inspiral and merger, as well as the
distributions of the remnant mass, spin, and recoil velocity.  For the
inspiral regime, we start with a random uniform distribution of spin
directions $\vec{S}_1$ and $\vec{S}_2$ over the sphere and magnitudes
$|\vec{S}_1/m_1^2|=|\vec{S}_2/m_2^2|=0.97$ for different mass ratios,
where
$\vec S_i$ and $m_i$ are the spin-angular momentum and mass of the
$i$th black hole.
Starting from a fiducial initial separation of $r_i=50M$, we perform
3.5-post-Newtonian-order evolutions down to a separation of $r_f=5M$, where
$M=m_1+m_2$, the total mass of the system. At this final fiducial
separation, we compute the angular distribution of the spins with
respect to the final orbital angular momentum, $\vec L$.
 We perform $16^4=65536$
simulations for six mass ratios between $q=1$ and $q=1/16$ and compute
the distribution of the angles $\hat{\vec{L}}\cdot\hat{\vec{\Delta}}$
and $\hat{\vec{L}}\cdot\hat{\vec{S}}$, directly related to recoil
velocities and total angular momentum. 
We find a small but statistically significant bias of the distribution
towards counter-alignment of both scalar products.  A post-Newtonian
analysis shows that radiation-reaction-driven dissipative effects on
the orbital angular momentum lead to this bias.
To study the merger of black-hole binaries, we turn to full numerical
techniques. In order to make use of the numerous simulations now
available in the literature, we introduce empirical formulae to
describe the final remnant black hole mass, spin, and recoil velocity
for merging black-hole binaries with arbitrary mass ratios and spins.
Our formulae are based on the post-Newtonian scaling, to model the
plunge phase, with amplitude
parameters chosen by a least-squares fit of recently available fully
nonlinear numerical simulations, supplemented by inspiral losses from
infinity to the ISCO. We then evaluate those formulae for
randomly chosen directions of the individual spins and magnitudes as
well as the binary's mass ratio. The number of evaluations has been
chosen such that there are 10 configurations per each dimension of
this parameter space, i.e. $10^7$. We found that the magnitude of the
recoil velocity distribution decays exponentially as $P(v) \sim \exp(-v/2500\
\KMS)$ with mean velocity $<v> = 630\ \kms$ and 
standard deviation $\sqrt{<v^2> - <v>^2} = 534\ \kms$, 
leading to a $23\%$ probability of recoils larger than $1000\
\KMS$, and a highly peaked angular distribution along
the final orbital axis. The studies of the distribution of the final
black-hole spin magnitude show a universal distribution highly peaked
at $S_f/m_f^2=0.73$ and a $25^\circ$ misalignment
with respect to the final orbital angular momentum, just prior to full
merger of the holes.
We also compute the statistical dependence of the magnitude of the recoil
velocity with respect to the ejection angle. The spin and recoil velocity 
distributions are also displayed as a function of the mass ratio. We finally
also compute the effects of the observer orientation with respect to the
recoil velocity vector to take into account the probabilities to measure a
given redshifted (or blueshifted) radial velocity of accretion disks with
respect to host galaxies.
\end{abstract}

 \maketitle

\section{Introduction}\label{sec:Introduction}

Astrophysical black-hole (BH) binaries are characterized by the mass ratio
$q=m_1/m_2 \leq 1$ of the smaller to larger BH, where $m_i$ is the
mass of BH $i$, the total mass $M=m_1+m_2$, eccentricity $e$ (assumed
to be very small), and spins $\vec S_i$ (where $S_i/m_i^2 < 1$). In
addition, it is often convenient to parameterize the binary with the
symmetric mass ratio $\eta = q/(1+q)^2$, specific spins $\vec
\alpha_i = \vec S_i/m_i^2$, total spin $\vec S = \vec S_1 + \vec S_2$,
$\vec \Delta = M (\vec S_2 /m_2 - \vec S_1/m_1)$,
 and orbital angular momentum $\vec L$.

The relevance of the spins to the dynamics of black-hole (BH) mergers
was recognized soon after the breakthrough in numerical relativity
\cite{Pretorius:2005gq, Campanelli:2005dd, Baker:2005vv} that allowed
for the long term, stable numerical evolution of such systems.
Notable examples of the early findings are the `hangup' effect
\cite{Campanelli:2006uy}, a repulsive spin-orbit interaction, that
delays the merger of black-hole binaries (BHB) when the spins are
aligned with the orbital angular momentum, and simultaneously causes
the system to radiate excess angular momentum, leading to a remnant BH
with sub-maximal spin.  The same mechanism produces an additional
attractive effect when the spins are counter-aligned with the orbital
angular momentum, leading to a prompt merger. Thus the radiation of
angular momentum and energy is asymmetric with respect to the relative
orientations of the total spin angular momentum vector and the orbital
angular momentum.

When the spins are not exactly aligned or counter aligned, new effects
appear (the hangup effect is still present). Precession of the spins
is important dynamically because it cause the orbital plane to
strongly precess just prior to merger~\cite{Campanelli:2006fy}.  The
final spin of the merged hole can flip with respect to the directions
of the individual ones, mainly due to the addition of the orbital
angular momentum \cite{Campanelli:2006fy}.  While the spin-orbit
coupling leads to strong precessional effects near merger, the
magnitudes of the spins are not affected to the same degree.  In
particular spin-orbit interactions are too weak to induce the binary
to corotate (or maintain corotation of an initially corotating binary)
at the last stages of the merger because the timescale for the
radiation driven inspiral is much smaller than the spin-orbit
interaction timescale~\cite{Campanelli:2006fg}.

Numerous other papers have studied different spin effects, such as the
large recoil velocities acquired by the remnant of the merger of two
spinning black holes~\cite{Herrmann:2007ac, Koppitz:2007ev,
Campanelli:2007ew, Baker:2007gi, Gonzalez:2007hi, Tichy:2007hk} and
long term evolutions of generic BHBs (i.e.\ unequal mass and unequal,
randomly-oriented spins)~\cite{Campanelli:2008nk, Szilagyi:2009qz}
 to cite a few of the
nearly one hundred papers published on the subject since 2006.

The characterization of the remnant black hole (BH) as the by-product
of a generic BH
binary (BHB) merger  is of great astrophysical interest as it allows
one to model the growth of BHs  during the evolution of the universe
and their effect on the dynamical evolution of galactic cores and
globular clusters, as well as the collisions of galaxies and stellar
size binary systems. Thanks to the recent breakthroughs in Numerical
Relativity~\citep{Pretorius:2005gq, Campanelli:2005dd, Baker:2005vv}
one can now precisely compute the masses, spins and recoil velocities
of these merged BHBs from fully nonlinear numerical simulations.

The modeling of the remnant black hole using fully-numerical
techniques was pioneered by the `Lazarus method' \citep{Baker:2003ds}
for spinning black holes followed by the breakthrough `moving
puncture' approach. In Refs.~\citep{Campanelli:2006uy,
Campanelli:2006fg, Campanelli:2006fy} the authors studied BHBs
characterized by equal-mass, equal-spin individual BHs, with the
spins aligned or counter-aligned with the orbital angular momentum,
using fully nonlinear numerical calculations and found a simple {\em
ad hoc} expression relating the final mass and spin of the remnant
with the spins of the individual BHs. This scenario was later
revisited in \citep{Rezzolla:2007xa, Rezzolla:2007rd} and the formula
for the remnant spin was generalized (by assuming that the angular
momentum is only radiated along the orbital axis, and neglecting the
energy loss) in \citep{Rezzolla:2007rz}
for arbitrary BH configurations (although in the latest paper of this
sequel this condition was removed \cite{Barausse:2009uz}.)
In \citep{Tichy:2008du} a more general
{\em ad hoc} fitting function was proposed. A more comprehensive
approach was proposed in \citep{Boyle:2007ru}; where a generic 
Taylor expansion,
reduced by the physical symmetries of the problem, was used to
fit the existing full numerical simulations. A
different approach was presented in \citep{Buonanno:2007sv} where
the particle limit approximation was extended to the equal-mass case
and the effects of post-ISCO (Innermost Stable Circular Orbit)
gravitational radiation were neglected.
This approach was further improved in~\citep{Kesden:2008ga} by taking
binding energies into account.  All of these approaches show a certain
degree of agreement with the remnant masses and spins obtained in the
few dozen fully nonlinear numerical simulations available, but
there remains significant uncertainties concerning their accuracy outside
this range of parameters.  In this paper we propose a set of formulae
that incorporate the benefits of both approaches in a unified way.

Due to the large astrophysical interest of computing remnant recoil
velocities, the modeling of recoil velocities followed an independent
path, particularly since the discovery \citep{Campanelli:2007ew,
Campanelli:2007cga}
that the spins of the black holes play a crucial role in producing
recoils of up to $4000\ \KMS$.
The importance of modeling the recoil velocities as a function of the
astrophysical parameters of the progenitor binary was quickly realized
\citep{Campanelli:2007ew, Baker:2007gi, Campanelli:2007cga}.

The news that the merger of binary black holes can produce recoil
velocities up to $4000\ \KMS$, and hence allow the remnant to escape from
major galaxies, led to numerous theoretical and observational efforts
to find traces of this phenomenon. Several studies made predictions of
specific observational features of recoiling supermassive black holes in the
cores of galaxies in the electromagnetic spectrum \citep{Haiman:2008zy,
Shields:2008va, Lippai:2008fx, Shields:2007ca, Komossa:2008ye,
Bonning:2007vt, Loeb:2007wz} from infrared \citep{Schnittman:2008ez} to
X-rays \citep{Devecchi:2008qy, Fujita:2008ka, Fujita:2008yi} and
morphological aspects of the galaxy cores \citep{Komossa:2008as,
Merritt:2008kg, Volonteri:2008gj}.  Notably, there began to appear
observations indicating the possibility of detection of such effects
\citep{Komossa:2008qd, Strateva:2008wt,Shields:2009jf}, and although alternative
explanations are possible \citep{Heckman:2008en, Shields:2008kn,
Bogdanovic:2008uz,Dotti:2008yb},  there is still the exciting possibility that these
observations can lead to the first confirmation of a prediction of
General Relativity in the highly-dynamical, strong-field regime.

In our approach to the recoil problem \citep{Campanelli:2007ew,
Campanelli:2007cga} we chose to use post-Newtonian theory as a guide
to model the recoil dependence on the physical parameters of the
progenitor BHB (See Eqs.
(3.31) in \citep{Kidder:1995zr}), while arguing that only full
numerical simulations can produce the correct amplitude of the effect.
Bearing this in mind, we proposed an empirical formula for the total
recoil velocities (see Eq.\ (\ref{eq:Pempirical}) below.) 
Our heuristic formula describing the recoil
velocity of a black-hole binary remnant as a function of the
parameters of the individual holes has been theoretically verified in
several ways. In~\cite{Campanelli:2007cga} the $\cos{\Theta}$
dependence was established and was confirmed in~\cite{Brugmann:2007zj}
for binaries with different initial separations. In
\cite{Herrmann:2007ex} the decomposition into spin components
perpendicular and parallel to the orbital plane was verified, and
in~\cite{Pollney:2007ss} it was found that the quadratic-in-spin
corrections to the in-plane recoil velocity are less than $20\ \KMS$.
Recently in \citep{Lousto:2008dn} we confirmed the leading $\eta^2$
(where $\eta = \frac{m_1 m_2}{(m_1+m_2)^2}$
 is the symmetric mass ratio) dependence of the large recoils
 out of the orbital plane.

Since the magnitude (and direction) of the recoil velocity of the remnant
black holes depend so sensitively on the spin orientation just around the
time of the formation of a common event horizon, it is important to 
establish that random oriented spins of individual black holes at large
separations (as a plausible initial astrophysical scenario) lead to randomly
oriented black holes near merger, or if there is some bias in their orientations
by the time they get very close together (i.e.\ at typical numerical 
simulations separations of a few $M$, where $M$ is the total mass). 
It has been argued recently \cite{Bogdanovic:2007hp,Perego:2009cw,Dotti:2009vz}
that the presence of gas and 
accretion of the individual black holes during the inspiral phase for
long time scales can lead to a preferential alignment of spins with the
orbital angular momentum, and hence to a configuration that leads to
modest recoil velocities (a few hundred $\KMS$). 
The latest results by Dotti et al \cite{Dotti:2009vz} indicate that
spin alignment occur in the scale of a few million years within 10 degrees
of the orbital angular momentum for cold disks and 30 degrees for warmer
disks. The proportion of wet to dry mergers in the
universe still needs to be established.
 Current rough estimates give comparable percentages for 
both kinds of mergers. With 3 to 5 mergers per galaxy during their lifetime
for current spiral and elliptic galaxies respectively
\footnote{M.Volonteri, private communication.}.
While the verification of these
claims for `wet' mergers is underway, in this paper we would like to
explore the possibility that such alignment (or counter-alignment) mechanism
exists for purely gravitational interactions (`dry mergers'). In general
we will seek to find if there is any bias in the individual spin distributions of black 
holes at close separations ($5-8M$ for starting typical full numerical
evolutions) when starting evolutions with post-Newtonian methods
at large  radii with  random spin orientations. These distributions, in turn, 
 will then help in choosing configuration for full numerical
simulations of close binaries.

The paper is organized as follows. In Section \ref{Sec:Inspiral} we describe
the post-Newtonian formalism to analyze the inspiral stage of the binary
evolutions. We use the Hamiltonian formulation (up to 3.5PN order)
to derive the equations of motion in the ADM-TT gauge.
Conservative and radiative
effects of the spins are included up to the next leading PN order. 
We also include a
purely analytic analysis of the projection of the quantity
$\vec\Delta$ along the orbital angular momentum $\vec L$,
which has a strong effect on the  recoil velocity, 
to qualitatively predict a slight bias towards counter-alignment of
these two vectors. The results of the statistics of numerical integration
of the post-Newtonian equations of motion (EOM) follows. We performed
integrations from initial separations of $r=50M$ with $16^4$ spin orientation chosen at random 
and magnitudes fixed at large astrophysical values, i.e. $S_i/m_i^2=0.97$
for different mass ratios in the range $1/16 \leq q=m_1/m_2\leq 1$. 
The results quantitatively confirm the bias towards counter-alignment of 
$\vec\Delta$ and total spin $\vec S$ with respect to the 
 orbital angular momentum $\vec L$.
Section \ref{Sec:Merger} deals with the merger phase, when the black holes
are much closer to each other and in a few orbits will merge into a single
larger one. This is the typical scenario that full numerical simulations assume.
The bulk properties of the remnant black hole can be summarized in terms
of empirical remnant formulae that describe its total mass, spin and recoil
velocity. We proposed formulae for these quantities based on post-Newtonian
scaling with amplitudes fixed by the full numerical simulations. With these
formulae at hand, we perform statistical studies by evaluation of these
expression for random distributions of mass ratios and individual spins
(magnitudes and spins). Those evaluations lead to a large recoil velocity
tail in the distribution with non negligible probabilities for
$v>1000\ \kms$,
and highly peaked about the direction of orbital angular momentum at
merger. Likewise,
evaluations of the final spin formulae lead to a wide distribution
peaked
at magnitudes of $S_f/M_f^2\approx0.73$ and orientations peaked at an angle
$\sim25^o$ with respect to the orbital angular momentum.
We complete the paper with a discussion of the astrophysical consequences
of these results and include an appendix with the computation of the 
innermost stable circular orbit radius, energy and angular momentum around 
Kerr black holes (needed for the remnant formulae), with an analytic 
solution for the equatorial and polar orbits.

\section{Inspiral phase of BHBs}\label{Sec:Inspiral}

\subsection{PN techniques}\label{SubSec:PN}

We construct the PN equations of motion using the formulae provided in
Refs.~\citep{Buonanno:2005xu, Damour:2007nc, Steinhoff:2007mb,
Steinhoff:2008ji}.
To obtain the conservative part of the PN equations of motion, we use
the following Hamiltonian,
\begin{eqnarray}
H &=& H_{\rm O,Newt} + H_{\rm O,1PN} + H_{\rm SO,1.5PN} 
+ H_{\rm O,2PN} 
\nonumber \\ &&
+ H_{\rm SS,2PN} + H_{\rm SO,2.5PN} + H_{\rm O,3PN} 
\nonumber \\ &&
+ H_{\rm S_1S_2,3PN} + H_{\rm S_1S_1(S_2S_2),3PN} 
\,,
\end{eqnarray}
where 
$H_{\rm O}$ contains the terms associated with the orbital motion
up to 3PN order, $H_{\rm
SO}$ contains the spin-orbit coupling terms up to 2.5PN order,
and $H_{\rm SS}$ contains the spin-spin coupling term up to 3PN order.
Note that Porto and Rothstein has discussed the spin-spin interaction 
by using effective field theory 
techniques~\cite{Porto:2006bt,Porto:2007tt,Porto:2008tb,Porto:2008jj}.
These are a very powerful approach to 
systematically discuss the dynamics of finite size objects. 

The equations of motion are then obtained via,
\begin{eqnarray}
\frac{dX^i}{dt} &=& \{ X^i , H \} = \frac{\partial H}{\partial P_i} 
\,, 
\label{eq:PNEOM1} \\
\frac{dP_i}{dt} &=& \{ P_i , H \} + F_i = - \frac{\partial H}{\partial X^i} + F_i 
\,, 
\label{eq:PNEOM2} \\
\frac{d \vec{S}_1}{dt}  &=& \{\vec{S}_1,H \} = \frac{\partial H}{\partial \vec{S}_1} \times \vec{S}_1 
\,, 
\label{eq:PNEOM3} \\
\frac{d \vec{S}_2}{dt}  &=& \{\vec{S}_2,H \} = \frac{\partial H}{\partial \vec{S}_2} \times \vec{S}_2 
\,, 
\label{eq:PNEOM4}
\end{eqnarray}
where $\{\cdot\cdot\cdot,\cdot\cdot\cdot\}$ denotes the Poisson brackets, 
$X^i = x_1^i - x_2^i$ and $P^i$ are relative coordinates and linear
momenta of the binary, $\vec{S}_1$ and $\vec{S}_2$ are the spins of each body,  
and $F_i$ is the radiation reaction force.
The radiation reaction force $\vec{F}$ is given by~\cite{Buonanno:2005xu},
\begin{eqnarray}
\vec{F} &=& \frac{1}{\omega \, \vert \vec{L} \vert} \, \frac{dE}{dt} \, \vec{P} 
\nonumber \\ && 
+ \frac{8}{15} \, \eta^2 \, \frac{v_{\omega}^8}{|\vec{L}|^2 R} \left\{
\left(61 + 48 \, \frac{m_2}{m_1} \right) \vec{P} \cdot \vec{S}_1 
\right. \nonumber \\ && \qquad \left. 
+ \left( 61 +48 \, \frac{m_1}{m_2} \right) \vec{P} \cdot \vec{S}_2 \right\} \vec{L} 
\,, 
\label{eq:RR}
\end{eqnarray}
where $\vec{L} = \vec{X}\times\vec{P}$, $R=|\vec{X}|$, 
$v_{\omega}=(M\omega)^{1/3}$ and $\omega$ is the orbital frequency. 
We use the following notation:
\begin{eqnarray}
M &=& m_1 + m_2 \,, \\
\delta M &=& m_1 - m_2 \,, \\
\eta &=& \frac{m_1m_2}{M^2} \,, \\
\vec{S} &=&  \vec{S}_1 + \vec{S}_2 \,, \\
\vec{\Delta} &=&  M\left(\frac{\vec{S}_2}{m_2} - \frac{\vec{S}_1}{m_1}\right) \,, \\
\vec{S}_0 &=& 2\vec{S} + {\frac{\delta m}{M}} \vec{\Delta}
\nonumber \\ 
&=& 
\left(1 + \frac{m_2}{m_1}\right)\vec{S}_1 + \left(1+\frac{m_1}{m_2}\right)\vec{S}_2 \,.  
\label{eq:Def_PN}
\end{eqnarray}
To calculate $dE/dt$, the instantaneous loss in energy, we use the
formulae given 
in Refs.~\cite{Arun:2008kb}
\footnote{We have used the corrected version of $dE/dt$ in
Appendix C of~\cite{Arun:2008kb}. 
Instead of Eq.~(C11) in~\cite{Arun:2008kb}, 
we need to use the 2.5PN order spin-orbit coupling effect 
in Eq.~(7.11) of \cite{Blanchet:2006gy} 
because the spin variables are defined with constant magnitude.  
In addition, in  Eq.~(C10) in ~\cite{Arun:2008kb}, 
one should remove the fourth term $\nu\{\cdot\cdot\cdot\}$ in the
expression (K.~G.~Arun, private communication). 
}.

\subsubsection{PN prediction of distribution of $\hat {\vec{L}}\cdot\hat {\vec{\Delta}}$}

The 
time derivative of the inner product 
$\hat {\vec{L}}\cdot\hat {\vec{\Delta}}$ 
where $\hat {\vec{L}}$ and $\hat {\vec{\Delta}}$ are the unit vector 
corresponding to $\vec{L}$ and $\vec{\Delta}$, respectively, is given
by
\begin{eqnarray}
\dot{(\hat {\vec{L}} \cdot \hat {\vec{\Delta}})} 
&=& \frac{\dot {\vec{L}} \cdot \vec{\Delta}}{|\vec{L}||\vec{\Delta}|} 
+ \frac{\vec{L} \cdot \dot {\vec{\Delta}}}{|\vec{L}||\vec{\Delta}|}
\nonumber \\ && 
- \frac{\vec{L} \cdot \vec{\Delta}\,|\vec{L}|^\cdot}{|\vec{L}|^2|\vec{\Delta}|} 
- \frac{\vec{L} \cdot \vec{\Delta}\,|\vec{\Delta}|^\cdot}{|\vec{L}||\vec{\Delta}|^2} 
\,.
\end{eqnarray}
Here since we focus only on the dissipative effect, 
we ignore $\dot {\vec{\Delta}}$ and $|\vec{\Delta}|^\cdot$ 
This is because there is no radiation reaction term 
in Eqs.~(\ref{eq:PNEOM3}) and (\ref{eq:PNEOM4}). 
The radiation reaction effect are introduced by the evolution equation of 
the linear momentum given in Eq.~(\ref{eq:PNEOM2}).  
Furthermore, we expect that the time evolution of the spin directions 
due to the conservative force will cancel out in a statistical treatment. 
Hence, we have 
\begin{eqnarray}
(\hat {\vec{L}} \cdot \hat {\vec{\Delta}})^{\cdot}_{\rm dis} 
&=& \frac{\dot {\vec{L}}_{\rm dis} \cdot \vec{\Delta}}{|\vec{L}||\vec{\Delta}|} 
- \frac{\vec{L} \cdot \vec{\Delta}\,|\vec{L}|^\cdot_{\rm dis}}{|\vec{L}|^2|\vec{\Delta}|} 
\,.
\end{eqnarray} 

The dissipative effect on the angular momentum is given by 
\begin{eqnarray}
\dot {\vec{L}}_{\rm dis} &=& \vec{X} \times \vec{F} 
\nonumber \\ 
&=& \frac{1}{\omega}\frac{dE}{dt} \hat {\vec{L}} 
\nonumber \\ && 
- \frac{8}{15} \, \eta^2 \, \frac{v_{\omega}^8\,R}{|\vec{L}|^2 } \left\{
\left(61 + 48 \, \frac{m_2}{m_1} \right) \vec{P} \cdot \vec{S}_1 
\right. \nonumber \\ && \left. 
+ \left( 61 +48 \, \frac{m_1}{m_2} \right) \vec{P} \cdot \vec{S}_2 \right\} \vec{P} 
\end{eqnarray}
where we have used the quasi-circular assumption and Eq.~(\ref{eq:RR}). 

Using this dissipation of the angular momentum, 
we obtain 
\begin{widetext}
\begin{eqnarray}
(\hat {\vec{L}} \cdot \hat {\vec{\Delta}})^{\cdot}_{\rm dis}  
&=& 
- \frac{8}{15} \, \frac{v_{\omega}^{11}}{M} 
\frac{q}{(1+q)^4} \, \frac{1}{|\vec{\alpha}_2-q \vec{\alpha}_1|} 
\biggl\{
- q^2\,\left(61\,q + 48 \right) (\hat {\vec{P}} \cdot \vec{\alpha}_1)^2 
+ \left( 61 +48 \, q \right) (\hat {\vec{P}} \cdot \vec{\alpha}_2)^2 
\nonumber \\ && \qquad 
+ q\,\left[
 \left(61\,q + 48 \right)  
- \left( 61 +48 \, q \right) 
\right] (\hat {\vec{P}} \cdot \vec{\alpha}_1)(\hat {\vec{P}} \cdot \vec{\alpha}_2)  
\biggr\} 
\,, 
\label{eq:PNpred}
\end{eqnarray}
\end{widetext}
in the leading PN order calculation. 
Here $\vec{\alpha}_1=\vec{S}_1/m_1^2$, $\vec{\alpha}_2=\vec{S}_2/m_2^2$ 
and $q=m_1/m_2$.  
Note that the $dE/dt$ term in $\dot {\vec{L}}_{\rm dis}$ cancels out,
and we have
\begin{widetext}
\begin{eqnarray}
(\hat {\vec{L}} \cdot \hat {\vec{S}})^{\cdot}_{\rm dis}  
&=& 
- \frac{8}{15} \, \frac{v_{\omega}^{11}}{M} 
\frac{q}{(1+q)^4} \, \frac{1}{|q^2 \vec{\alpha}_1 + \vec{\alpha}_2|} 
\biggl\{
q^3\,\left(61\,q + 48 \right) (\hat {\vec{P}} \cdot \vec{\alpha}_1)^2 
+ \left( 61 +48 \, q \right) (\hat {\vec{P}} \cdot \vec{\alpha}_2)^2 
\nonumber \\ && \qquad 
+ q\,\left[
 \left(61\,q + 48 \right)  
+ q\,\left( 61 +48 \, q \right) 
\right] (\hat {\vec{P}} \cdot \vec{\alpha}_1)(\hat {\vec{P}} \cdot \vec{\alpha}_2)  
\biggr\} 
\,, 
\label{eq:PNpredS}
\end{eqnarray}
\end{widetext}

Next, we consider the time integration from $t=t_i$ to $t=t_f$.  
\begin{widetext}
\begin{eqnarray}
\int_{t_i}^{t_f} (\hat {\vec{L}} \cdot \hat {\vec{\Delta}})^{\cdot}_{\rm dis} dt
&=& -\frac{5}{64} \frac{(1+q)^2}{q} \int_{R_i}^{R_f} 
(\hat {\vec{L}} \cdot \hat {\vec{\Delta}})^{\cdot}_{\rm dis} \left(\frac{M}{R}\right)^{-3} dR 
\nonumber \\ 
&=& -\frac{1}{36} 
\frac{1}{(1+q)^2} \, \frac{1}{|\vec{\alpha}_2-q \vec{\alpha}_1|} \biggl\{
- q^2\,\left(61\,q + 48 \right) (\hat {\vec{P}} \cdot \vec{\alpha}_1)^2 
+ \left( 61 +48 \, q \right) (\hat {\vec{P}} \cdot \vec{\alpha}_2)^2 
\nonumber \\ && \qquad 
+ q\,\left[
 \left(61\,q + 48 \right)  
- \left( 61 +48 \, q \right) 
\right] (\hat {\vec{P}} \cdot \vec{\alpha}_1)(\hat {\vec{P}} \cdot \vec{\alpha}_2)  
\biggr\}
\left[ 
\left(\frac{M}{R_f}\right)^{3/2}-\left(\frac{M}{R_i}\right)^{3/2}
\right] \,,
\nonumber \\ 
\int_{t_i}^{t_f} (\hat {\vec{L}} \cdot \hat {\vec{S}})^{\cdot}_{\rm dis} dt
&=& -\frac{5}{64} \frac{(1+q)^2}{q} \int_{R_i}^{R_f} 
(\hat {\vec{L}} \cdot \hat {\vec{S}})^{\cdot}_{\rm dis} \left(\frac{M}{R}\right)^{-3} dR 
\nonumber \\ 
&=& -\frac{1}{36} 
\frac{1}{(1+q)^2} \, \frac{1}{|q^2 \vec{\alpha}_1 + \vec{\alpha}_2|} 
\biggl\{
q^3\,\left(61\,q + 48 \right) (\hat {\vec{P}} \cdot \vec{\alpha}_1)^2 
+ \left( 61 +48 \, q \right) (\hat {\vec{P}} \cdot \vec{\alpha}_2)^2 
\nonumber \\ && \qquad 
+ q\,\left[
 \left(61\,q + 48 \right)  
+ q\,\left( 61 +48 \, q \right) 
\right] (\hat {\vec{P}} \cdot \vec{\alpha}_1)(\hat {\vec{P}} \cdot \vec{\alpha}_2)  
\biggr\} 
\left[ 
\left(\frac{M}{R_f}\right)^{3/2}-\left(\frac{M}{R_i}\right)^{3/2}
\right] \,,
\label{eq:IntF}
\end{eqnarray}
\end{widetext}
where we considered only the evolution of $v_{\omega}$, i.e., the inspiral, 
and have used the leading radiation reaction and the Newtonian velocity, 
\begin{eqnarray}
\frac{dR}{dt} &=& -\frac{64}{5} \frac{q}{(1+q)^2} \left(\frac{M}{R}\right)^3 \,,
\nonumber \\ 
v_{\omega} &=& \sqrt{\frac{M}{R}} \,.
\end{eqnarray}

In the above integration, we derived the formula 
assuming $\hat {\vec{P}} \cdot \vec{\alpha}_i=$ 
constant ($i=1,\,2$). However, since the spins precess,  
we need to check the evolution of $\hat {\vec{P}} \cdot \vec{\alpha}_i$. 
From the evolution equations for spins in the leading PN order, 
the evolution equations 
for $\hat {\vec{P}} \cdot \vec{\alpha}_i$ are given by
\begin{eqnarray}
\left( \hat {\vec{P}} \cdot \vec{\alpha}_1 \right)^{\cdot} 
&=&  
\left[ -\frac{v_{\omega}}{R} 
+ \frac{M v_{\omega}}{R^2} \frac{1}{(1+q)^2} \left(2q+\frac{3}{2} \right) \right] 
\nonumber \\ && \times 
\left( \hat {\vec{X}} \cdot \vec{\alpha}_1 \right) 
\,, 
\nonumber \\ 
\left( \hat {\vec{P}} \cdot \vec{\alpha}_2 \right)^{\cdot} 
&=&  
\left[ -\frac{v_{\omega}}{R} 
+ \frac{M v_{\omega}}{R^2} \frac{q}{(1+q)^2} \left(2+\frac{3}{2}q \right) \right] 
\nonumber \\ && \times 
\left( \hat {\vec{X}} \cdot \vec{\alpha}_2 \right) 
\,, 
\end{eqnarray}

We note that in the limit $q \to 0$, we only need
to consider the evolution of $\hat {\vec{P}} \cdot \vec{\alpha}_2$ 
in Eq.~(\ref{eq:PNpred}), 
\begin{eqnarray}
\dot{\left( \hat {\vec{P}} \cdot \vec{\alpha}_2 \right)} 
&=& - \frac{v_{\omega}}{R} 
\left( \hat {\vec{X}} \cdot \vec{\alpha}_2 \right) \,. 
\end{eqnarray}
This equation means that the direction of $\vec{\alpha}_2$ does not change, 
i.e., there is no precession of the spin. 
Therefore, we may replace $(\hat {\vec{P}} \cdot \vec{\alpha}_2)^2$
in Eq.~(\ref{eq:IntF}) by the one-orbit average 
$<(\hat {\vec{P}} \cdot \vec{\alpha}_2)^2>_t$ of $(\hat {\vec{P}} \cdot \vec{\alpha}_2)^2$.
Although the adiabatic evolution of $<(\hat {\vec{P}} \cdot
\vec{\alpha}_2)^2>_t$  is present,
its effect comes in at higher PN order in Eq.~(\ref{eq:IntF}). 
In this case, it should be noted that we may consider a test particle 
orbiting around a Kerr black hole with the spin $\vec{S}_2$. 
According to~\cite{Ganz:2007rf} in the black hole perturbation approach, 
the particle's angular momentum 
and the black hole's spin tend to be anti-parallel. 

On the other hand, in the case of comparable mass binaries, 
the direction of $\vec{\alpha}_i$ changes on a timescale much shorter
than the integration time. 
Hence, Eq.~(\ref{eq:IntF}) is not expected to be accurate in
$q\to1$ limit. 

In Table~\ref{table:PNP}, we show the $q$ dependence of Eq.~(\ref{eq:IntF}) 
when we ignore the spin precession. 
Here, we take the average with respect to the direction of two spins  
to represent the randomly oriented spins. 
We also present the spin amplitude dependence in Table~\ref{table:PNPchi}. 

\begin{table}
\caption{The $q$ dependence in the evolution of $(\hat {\vec{L}} \cdot \hat {\vec{\Delta}})_{\rm dis}$ 
and $(\hat {\vec{L}} \cdot \hat {\vec{S}})_{\rm dis}$ from $r=50M$ to $r=5M$. 
We set $|\vec{\alpha}_1| = |\vec{\alpha}_2| = 0.97$. }
\label{table:PNP}
\begin{ruledtabular}
\begin{tabular}{lll}
$q$ & $(\hat {\vec{L}} \cdot \hat {\vec{\Delta}})_{\rm dis}$ 
& $(\hat {\vec{L}} \cdot \hat {\vec{S}})_{\rm dis}$ \\
\hline
1.00   & $0.0000$  & $-0.0283$ \\ 
0.75   & $-0.0111$ & $-0.0287$ \\
0.50   & $-0.0224$ & $-0.0310$ \\
0.25   & $-0.0343$ & $-0.0366$ \\
0.125  & $-0.0406$ & $-0.0412$ \\
0.0625 & $-0.0440$ & $-0.0441$ \\
0.00   & $-0.0475$ & $-0.0475$ \\
\end{tabular}
\end{ruledtabular}
\end{table}

\begin{table}
\caption{The amplitude dependence of the spin in the evolution 
of $(\hat {\vec{L}} \cdot \hat {\vec{\Delta}})_{\rm dis}$ 
and $(\hat {\vec{L}} \cdot \hat {\vec{S}})_{\rm dis}$ from $r=50M$ to $r=5M$. 
We set $q=0.25$ and $|\vec{\alpha}_1| = |\vec{\alpha}_2| = \alpha$. }
\label{table:PNPchi}
\begin{ruledtabular}
\begin{tabular}{lll}
$\alpha$ & $(\hat {\vec{L}} \cdot \hat {\vec{\Delta}})_{\rm dis}$ 
& $(\hat {\vec{L}} \cdot \hat {\vec{S}})_{\rm dis}$ \\
\hline
0.97          & $-0.0343$ & $-0.0366$ \\ 
$0.97/\sqrt{2}$ & $-0.0242$ & $-0.0259$ \\
0.97/2        & $-0.0171$ & $-0.0183$ \\
0.97/4        & $-0.0086$ & $-0.0092$ \\
0.97/8        & $-0.0043$ & $-0.0046$ \\
0.97/16       & $-0.0021$ & $-0.0023$ \\
\end{tabular}
\end{ruledtabular}
\end{table}

\subsection{Statistical Results}\label{SubSec:StatsPN}

For our PN evolutions with used an adaptive fourth-order Runge-Kutta
time-integration scheme with a relative tolerance of $10^{-13}$. The
initial data for the simulations were generated using the 3PN
conservative equations for quasi-circular orbits with orbital
frequency $M\Omega = 0.00275$, which corresponds to an orbital radius
of $50\pm2M$. In most cases we stopped the PN simulations at a fixed
orbital radius of $5M$, but also performed a set of simulations that
terminated at $r=8M$ in order to see the effect of the final orbital
radius on the distributions.  To obtain the initial PN orbital
parameters, we used uniform distributions of $\vec \alpha_1$ and $\vec
\alpha_2$ over the sphere (by choosing uniform random distributions in
$\mu = \cos \theta$ and $\phi$) with fix amplitude $\alpha = 0.97$. We
produced 65536 random spin configurations for each fix mass ratio
$q=1, 3/4, 1/2, 1/4, 1/8, 1/16$. Each run took approximately 10
minutes. In addition we performed  sets of 65536 run for $q=1/4$ and
$\alpha = 0.97/\sqrt{2}$, $\alpha=0.97/2$, as well as
$\alpha=0.97/\sqrt{2}$ but terminating at $r=8M$ rather than $r=5M$.
We denote these three latter distributions in Table~\ref{table:fits}
by 0.25S1, 0.25S2, and 0.25F, respectively.

\begin{figure}
\includegraphics[width=3.5in]{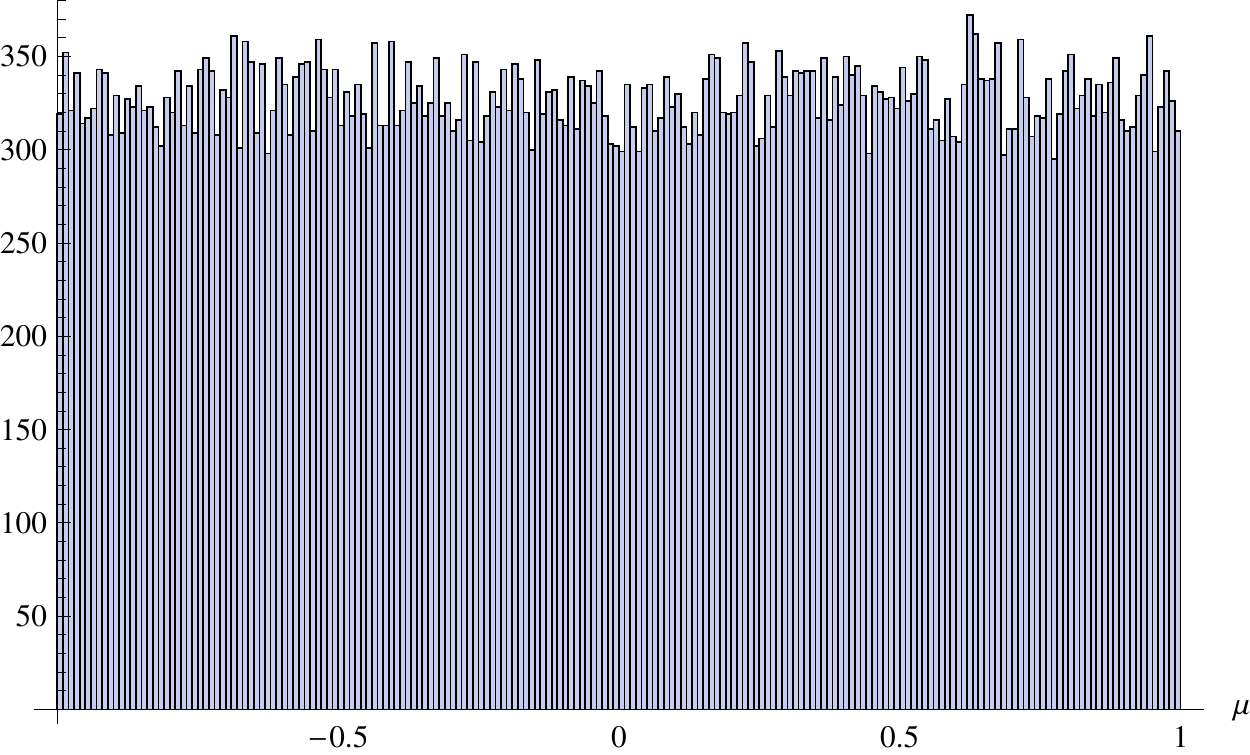}
\caption{The $P(\mu = \hat{\vec L} \cdot \hat{\vec\Delta})$
distribution for $q=1$ at $r=5M$ starting from a uniform distribution
at $r=50M$. Here we plot the number of events in the given range of
$\mu$ out of $16^4$ total events. }
\label{fig:q1.00dist}
\end{figure}

\begin{figure}
\includegraphics[width=3.5in]{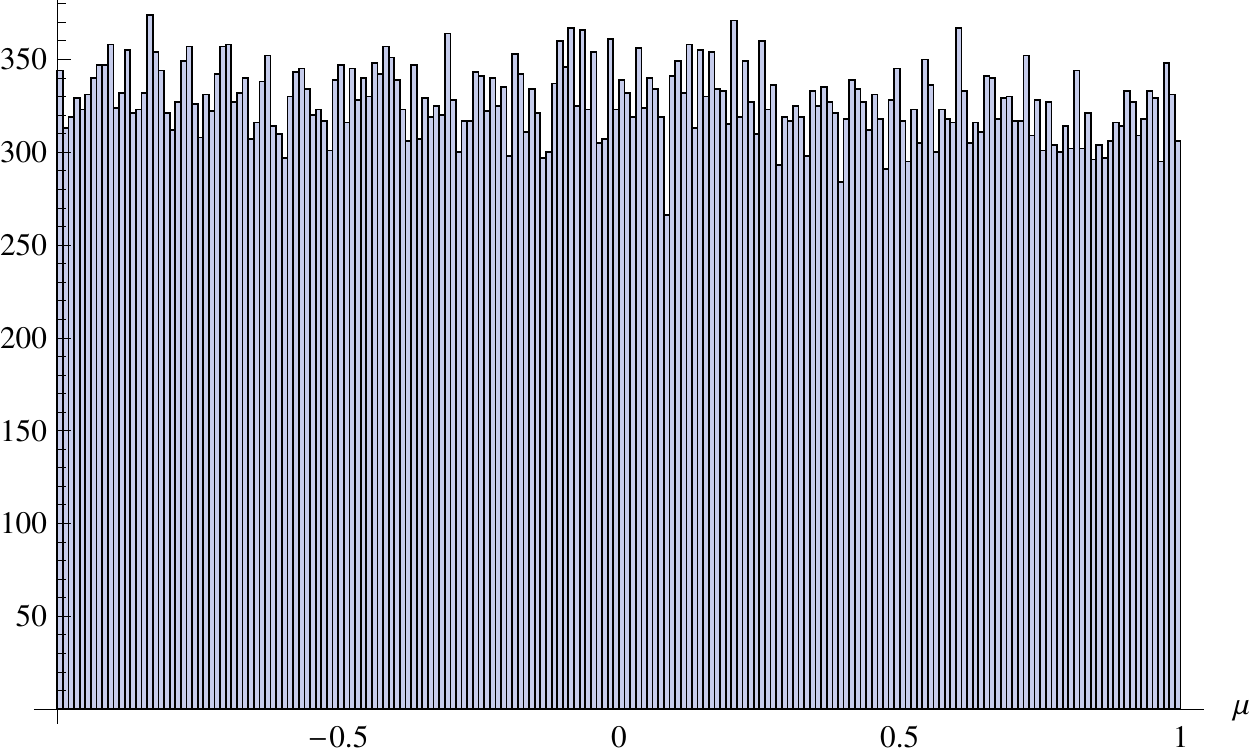}
\caption{The $P(\mu = \hat{\vec L} \cdot \hat{\vec\Delta})$
distribution for $q=3/4$ at $r=5M$ starting from a uniform distribution
at $r=50M$. Here we plot the number of events in the given range of
$\mu$ out of $16^4$ total events. }
\label{fig:q0.75dist}
\end{figure}

\begin{figure}
\includegraphics[width=3.5in]{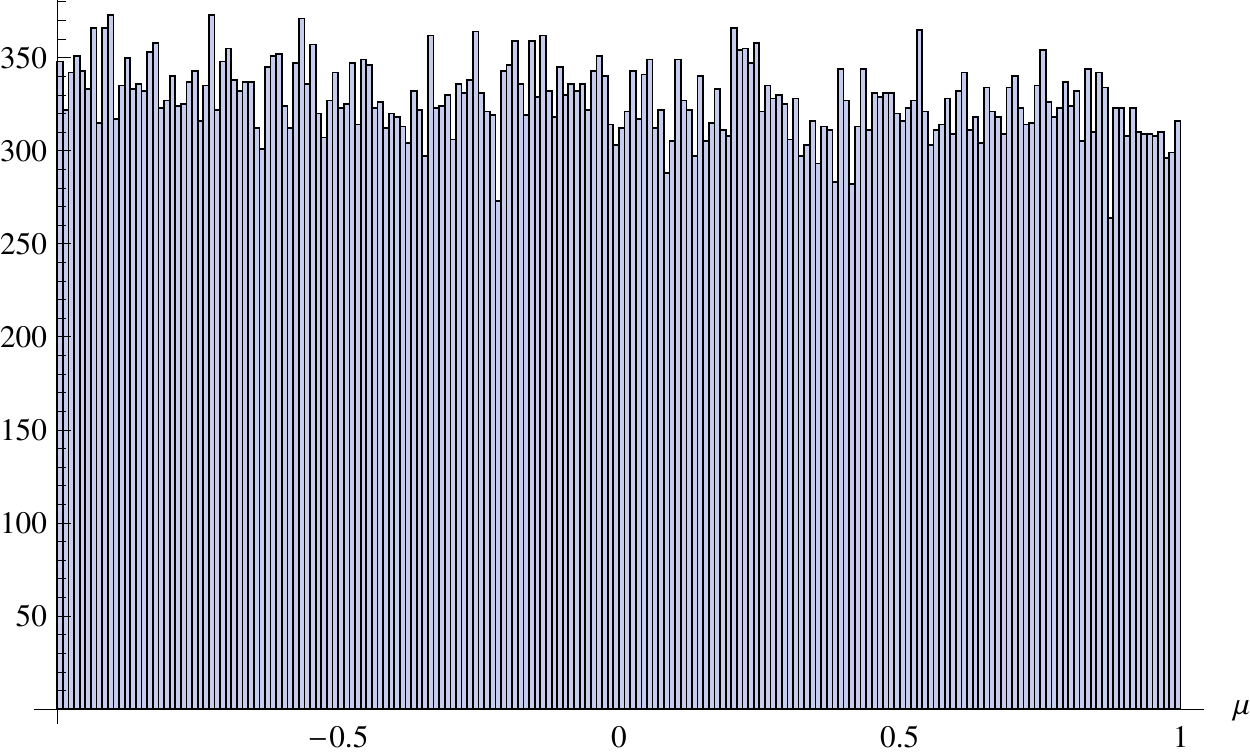}
\caption{The $P(\mu = \hat{\vec L} \cdot \hat{\vec\Delta})$
distribution for $q=1/2$ at $r=5M$ starting from a uniform distribution
at $r=50M$. Here we plot the number of events in the given range of
$\mu$ out of $16^4$ total events. }
\label{fig:q0.50dist}
\end{figure}

\begin{figure}
\includegraphics[width=3.5in]{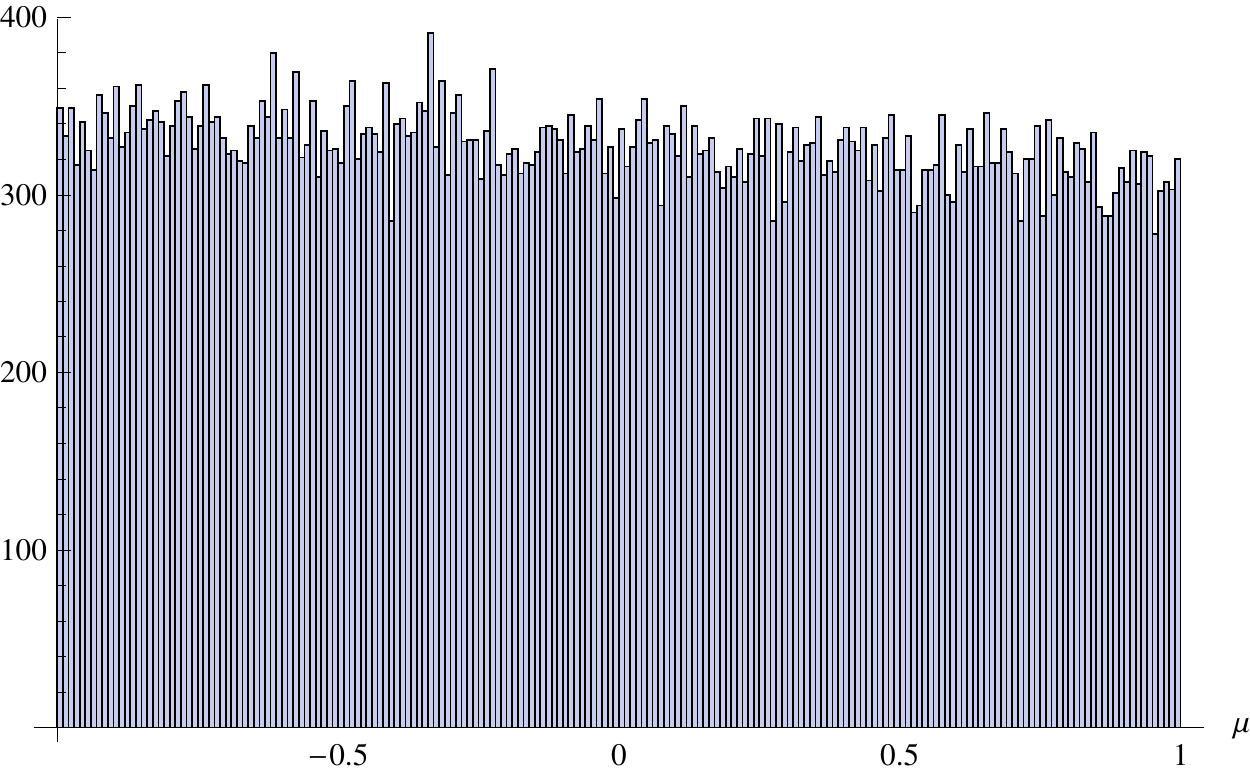}
\caption{The $P(\mu = \hat{\vec L} \cdot \hat{\vec\Delta})$
distribution for $q=1/4$ at $r=5M$ starting from a uniform distribution
at $r=50M$. Here we plot the number of events in the given range of
$\mu$ out of $16^4$ total events. }
\label{fig:q0.25dist}
\end{figure}

\begin{figure}
\includegraphics[width=3.5in]{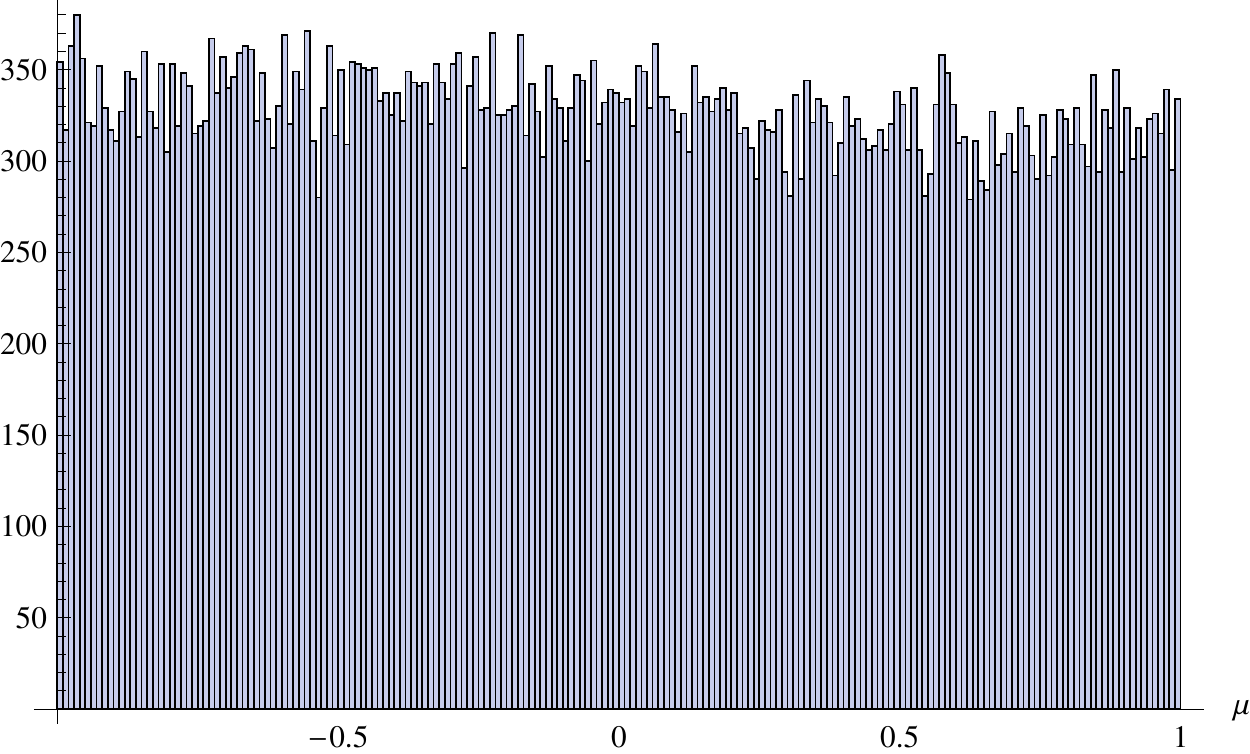}
\caption{The $P(\mu = \hat{\vec L} \cdot \hat{\vec\Delta})$
distribution for $q=1/8$ at $r=5M$ starting from a uniform distribution
at $r=50M$. Here we plot the number of events in the given range of
$\mu$ out of $16^4$ total events. }
\label{fig:q0.125dist}
\end{figure}

\begin{figure}
\includegraphics[width=3.5in]{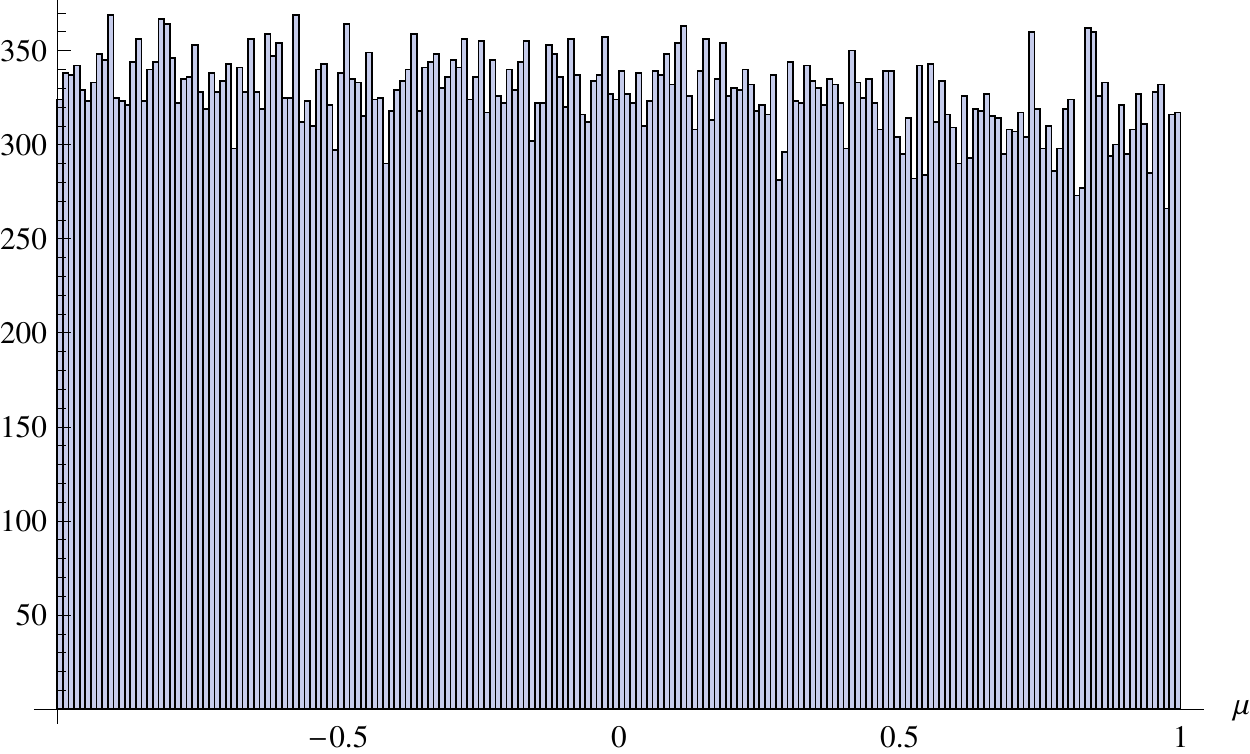}
\caption{The $P(\mu = \hat{\vec L} \cdot \hat{\vec\Delta})$
distribution for $q=1/16$ at $r=5M$ starting from a uniform distribution
at $r=50M$. Here we plot the number of events in the given range of
$\mu$ out of $16^4$ total events. }
\label{fig:q0.062dist}
\end{figure}

\begin{figure}
\includegraphics[width=3.5in]{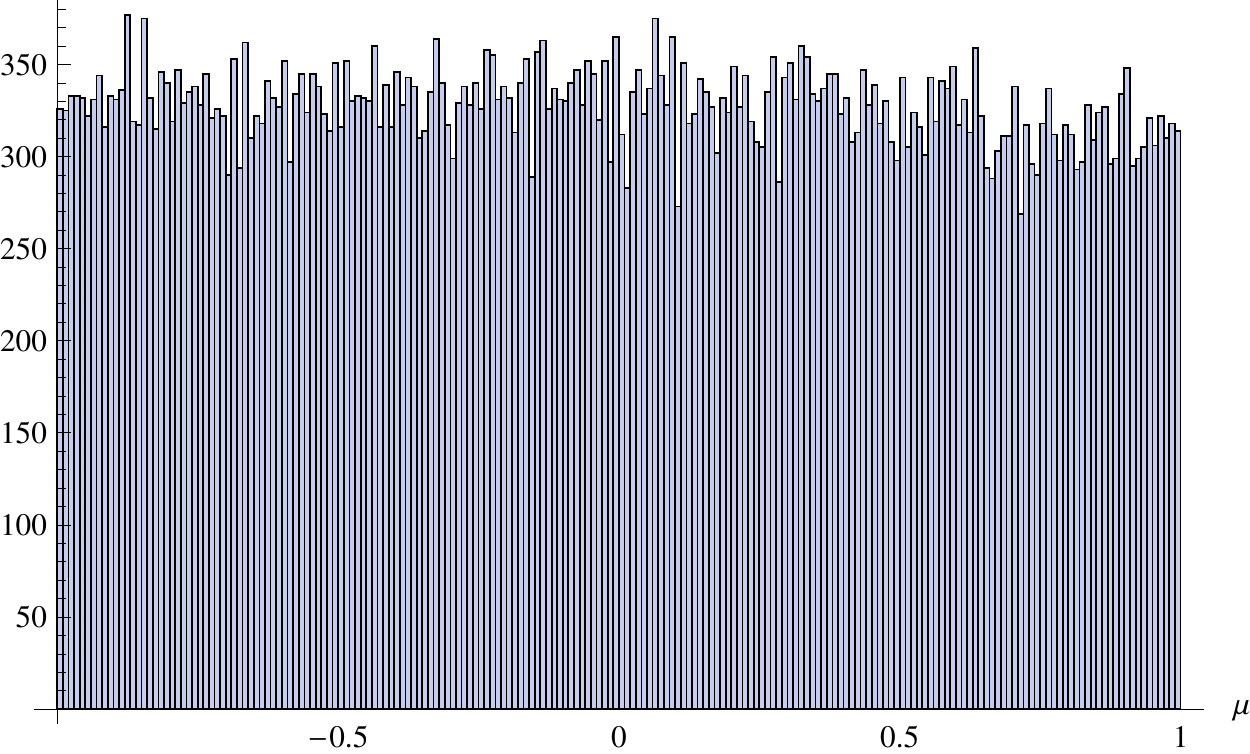}
\caption{The $P(\mu = \hat{\vec L} \cdot \hat{\vec\Delta})$
distribution for $q=1/4$ at $r=5M$ starting from a uniform distribution
at $r=50M$ and $\alpha_1 = \alpha_2 = 0.97/\sqrt{2}$. Here we plot the number of events in the given range of
$\mu$ out of $16^4$ total events. }
\label{fig:q0.25smalldist}
\end{figure}

\begin{figure}
\includegraphics[width=3.5in]{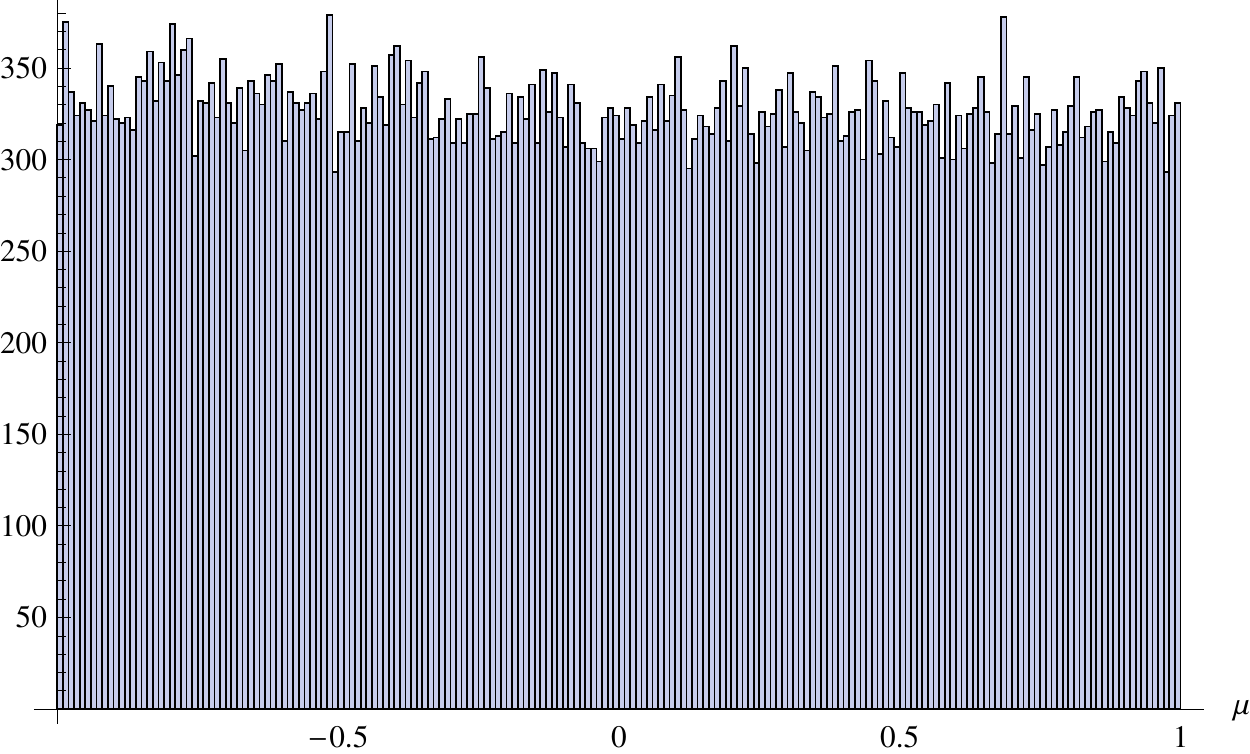}
\caption{The $P(\mu = \hat{\vec L} \cdot \hat{\vec\Delta})$
distribution for $q=1/4$ at $r=5M$ starting from a uniform distribution
at $r=50M$ and $\alpha_1 = \alpha_2 = 0.97/2$. Here we plot the number of events in the given range of
$\mu$ out of $16^4$ total events. }
\label{fig:q0.25vsmalldist}
\end{figure}

\begin{figure}
\includegraphics[width=3.5in]{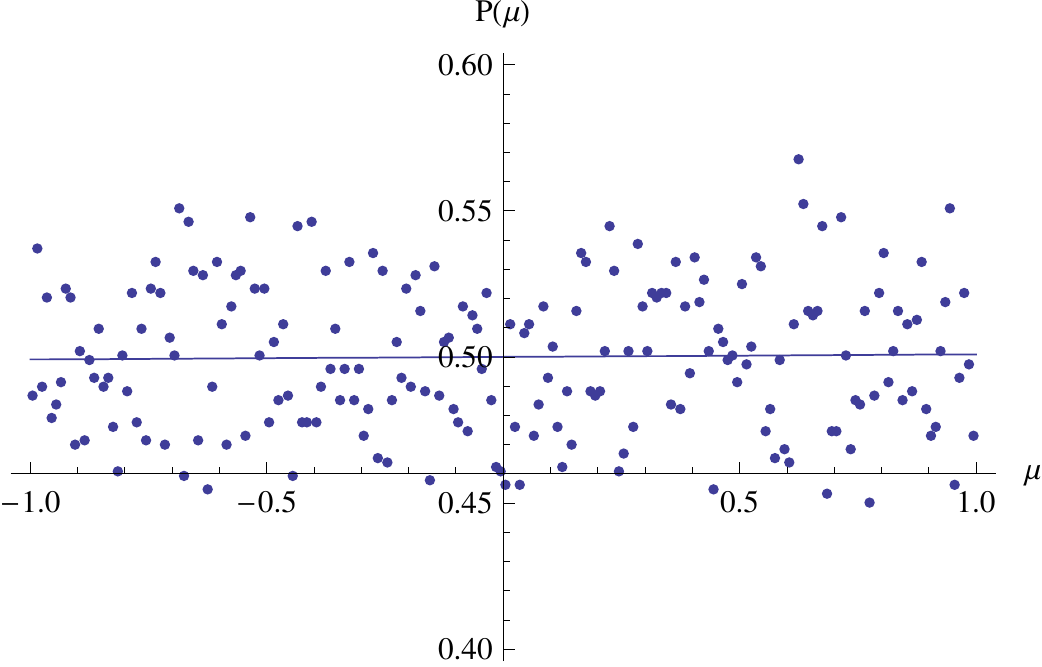}
\caption{The fit to the normalized 
$P(\mu = \hat{\vec L} \cdot \hat{\vec\Delta})$ distribution at $r=5M$
for $q=1$. The data have been binned with a bin width of $\delta\mu =
0.01$ and normalized to a total probability of 1.}
\label{fig:q1.00fit}
\end{figure}

\begin{figure}
\includegraphics[width=3.5in]{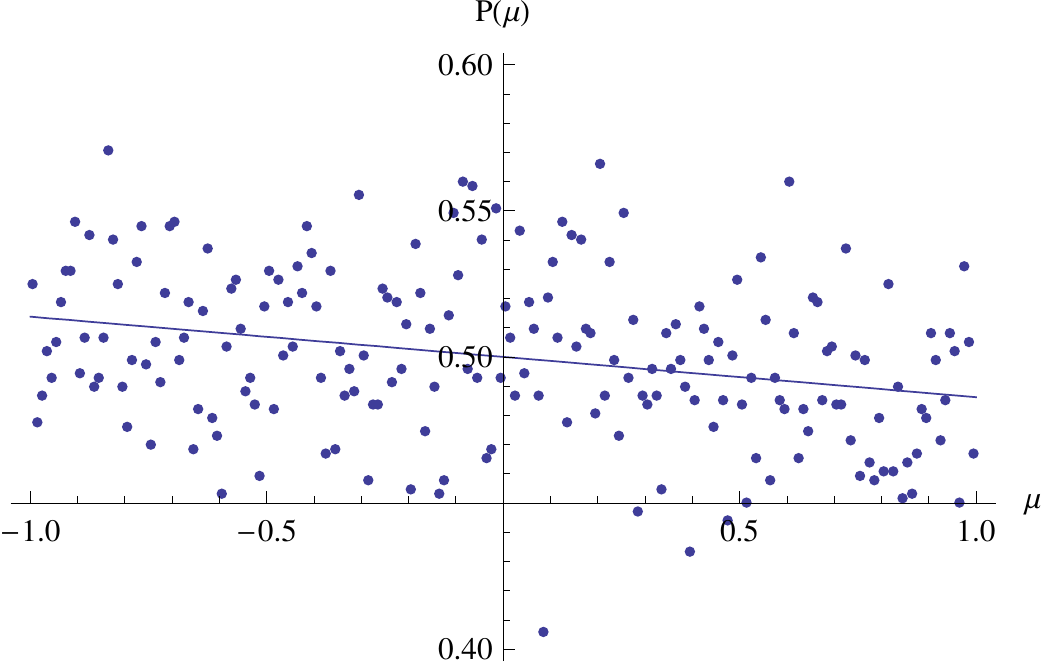}
\caption{The fit to the normalized 
$P(\mu = \hat{\vec L} \cdot \hat{\vec\Delta})$ distribution at $r=5M$
for $q=3/4$. The data have been binned with a bin width of $\delta\mu =
0.01$ and normalized to a total probability of 1.}
\label{fig:q0.75fit}
\end{figure}

\begin{figure}
\includegraphics[width=3.5in]{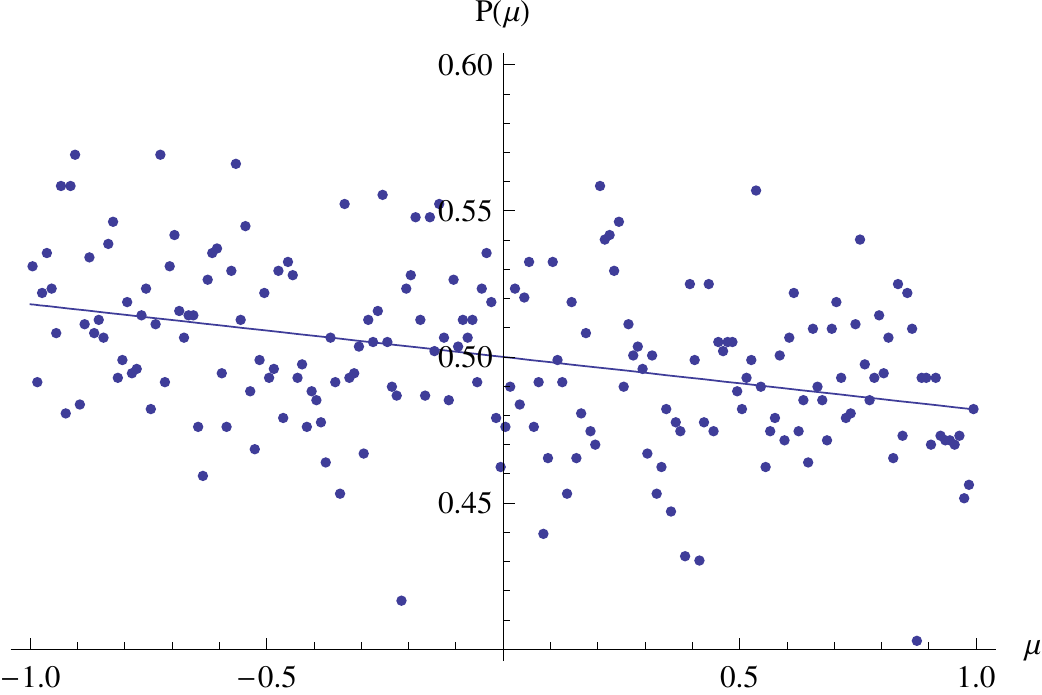}
\caption{The fit to the normalized 
$P(\mu = \hat{\vec L} \cdot \hat{\vec\Delta})$ distribution at $r=5M$
for $q=1/2$. The data have been binned with a bin width of $\delta\mu =
0.01$}
\label{fig:q0.50fit}
\end{figure}

\begin{figure}
\includegraphics[width=3.5in]{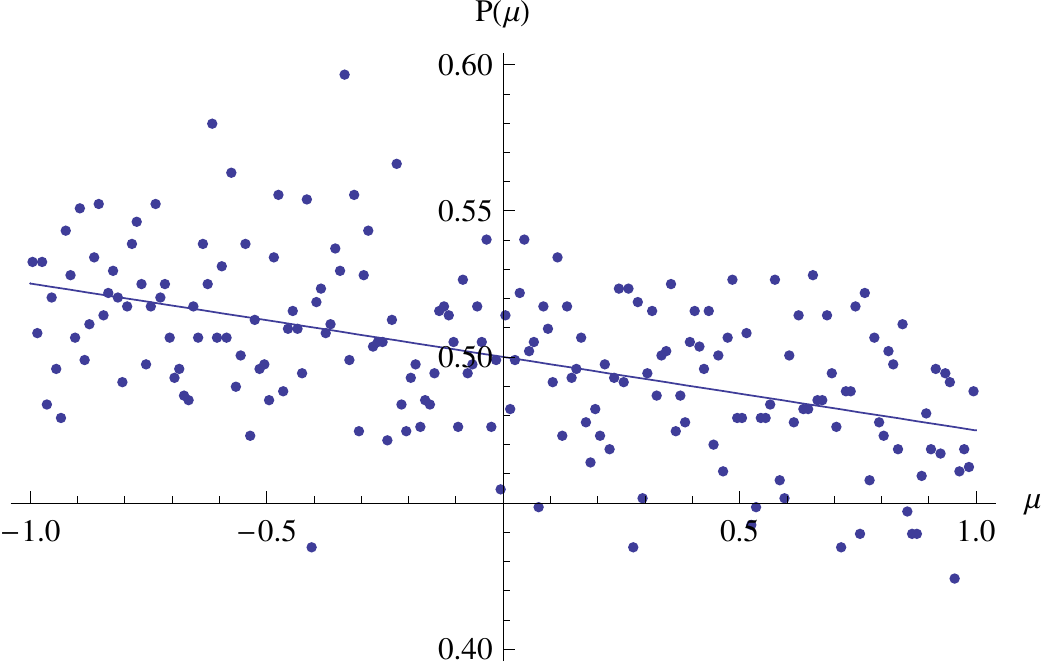}
\caption{The fit to the normalized 
$P(\mu = \hat{\vec L} \cdot \hat{\vec\Delta})$ distribution at $r=5M$
for $q=1/4$. The data have been binned with a bin width of $\delta\mu =
0.01$ and normalized to a total probability of 1.}
\label{fig:q0.25fit}
\end{figure}

\begin{figure}
\includegraphics[width=3.5in]{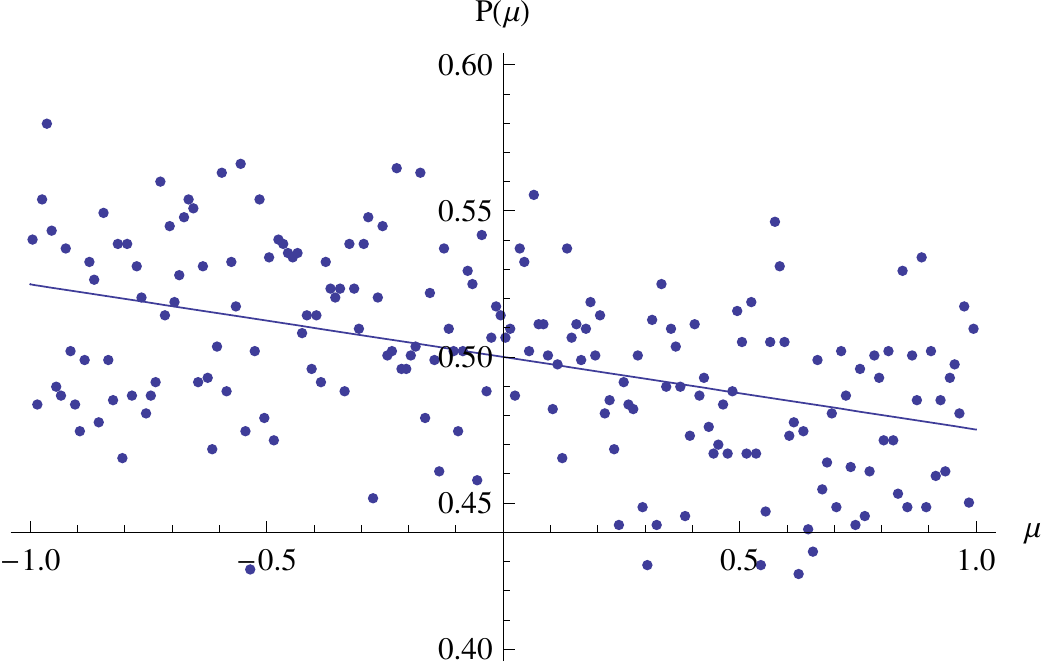}
\caption{The fit to the normalized 
$P(\mu = \hat{\vec L} \cdot \hat{\vec\Delta})$ distribution at $r=5M$
for $q=1/8$. The data have been binned with a bin width of $\delta\mu =
0.01$ and normalized to a total probability of 1.}
\label{fig:q0.125fit}
\end{figure}

\begin{figure}
\includegraphics[width=3.5in]{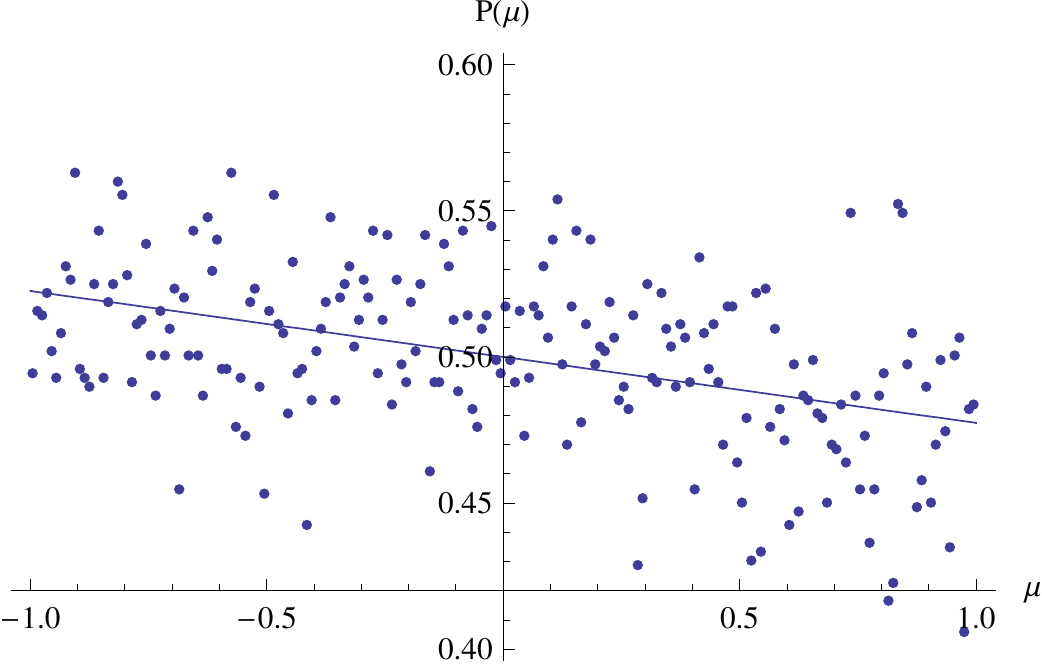}
\caption{The fit to the normalized 
$P(\mu = \hat{\vec L} \cdot \hat{\vec\Delta})$ distribution at $r=5M$
for $q=1/16$. The data have been binned with a bin width of $\delta\mu =
0.01$ and normalized to a total probability of 1.}
\label{fig:q0.062fit}
\end{figure}

\begin{figure}
\includegraphics[width=3.5in]{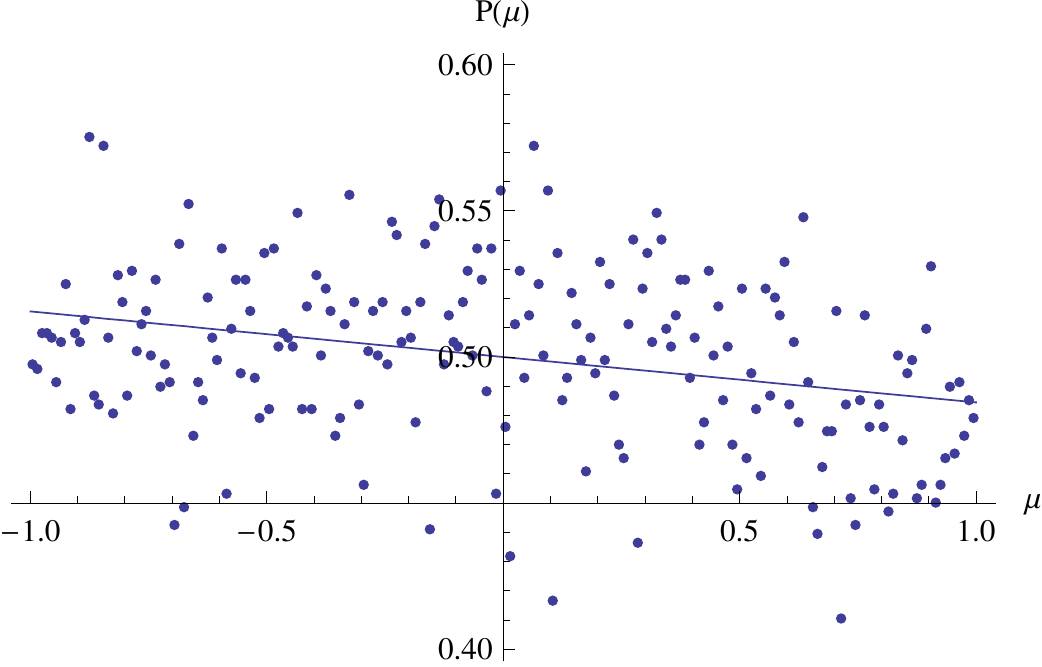}
\caption{The fit to the normalized 
$P(\mu = \hat{\vec L} \cdot \hat{\vec\Delta})$ distribution at $r=5M$
for $q=1/4$ and $\alpha_1 = \alpha_2 = 0.97/\sqrt{2}$. The data have been binned with a bin width of $\delta\mu =
0.01$ and normalized to a total probability of 1.}
\label{fig:q0.25smallfit}
\end{figure}

\begin{figure}
\includegraphics[width=3.5in]{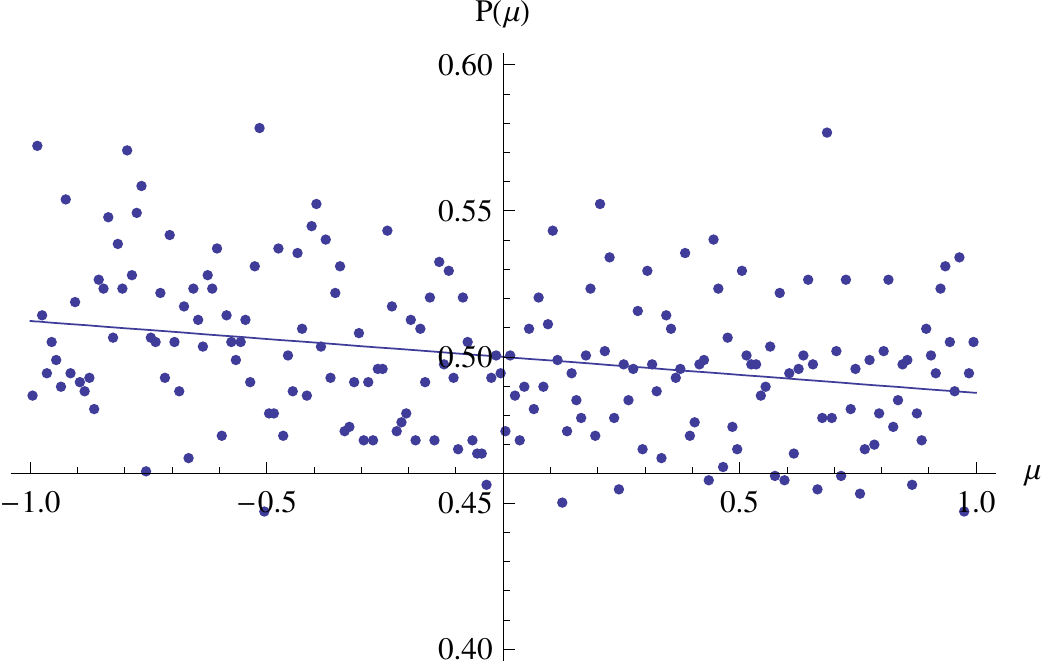}
\caption{The fit to the normalized 
$P(\mu = \hat{\vec L} \cdot \hat{\vec\Delta})$ distribution at $r=5M$
for $q=1/4$ and $\alpha_1 = \alpha_2 = 0.97/2$. The data have been binned with a bin width of $\delta\mu =
0.01$ and normalized to a total probability of 1.}
\label{fig:q0.25vsmallfit}
\end{figure}

\begin{figure}
\includegraphics[width=3.5in]{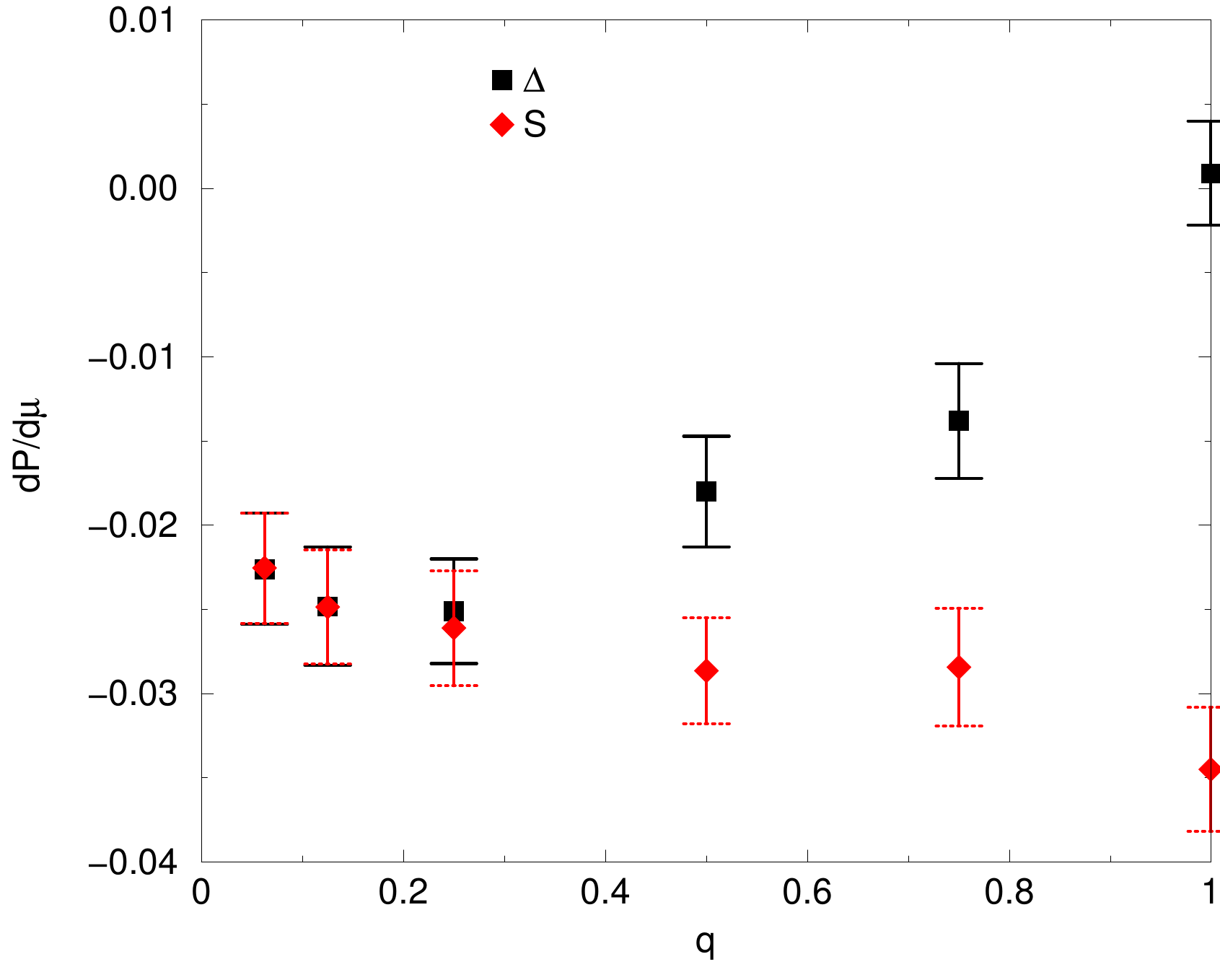}
\caption{The dependence of the slope in the distribution of the angle
between $\vec \Delta$ and the orbital angular momentum, as well as the
angle between $\vec S$ and the orbital angular momentum
as a function of mass ratio.}
\label{fig:dpdmu}
\end{figure}

\begin{figure}
\includegraphics[width=3.5in]{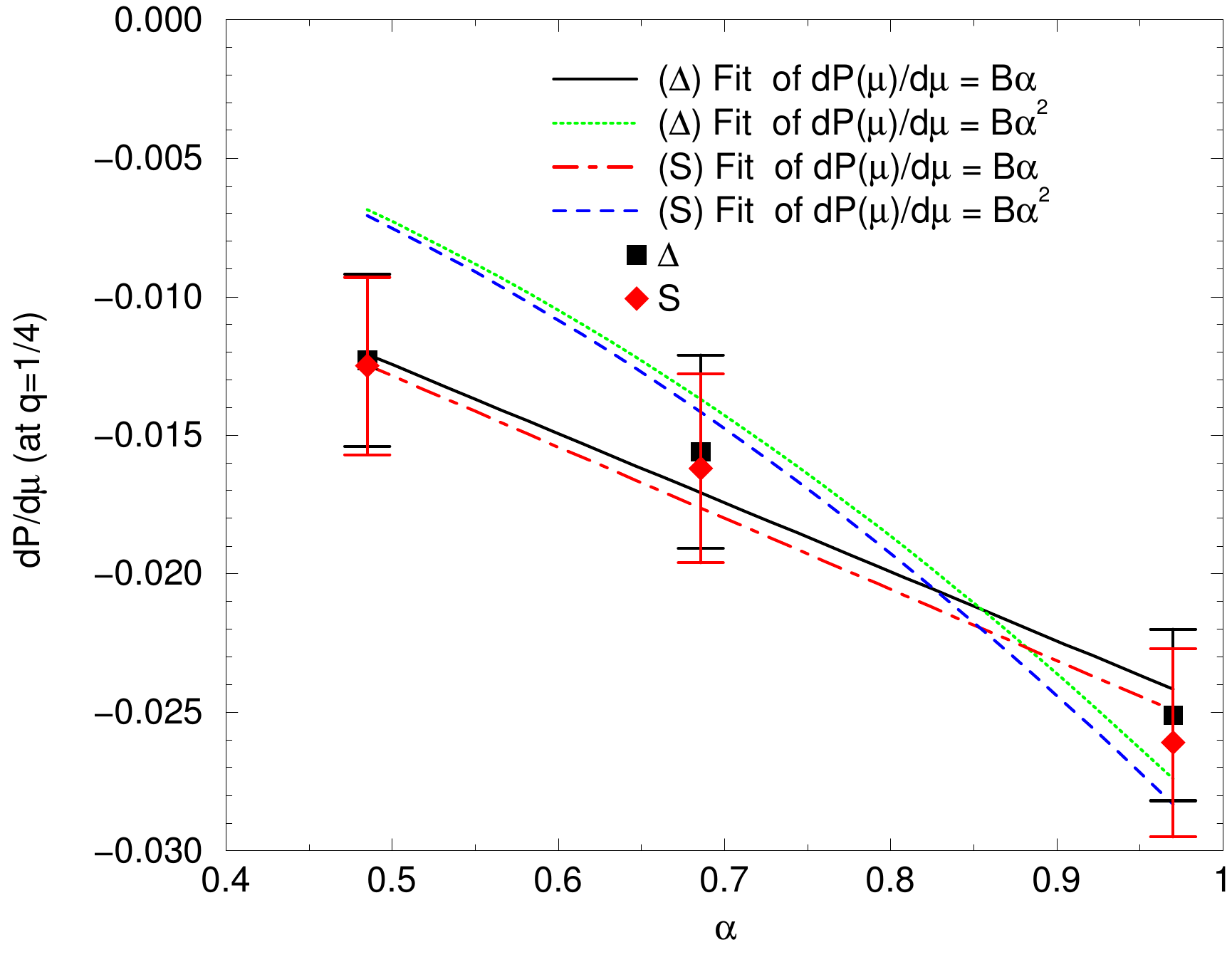}
\caption{The dependence of the slope in the distribution of the angle
between $\vec \Delta$ and the orbital angular momentum $\vec L$ for $q=1/4$ as
a function of $|\vec\alpha_1| = |\vec \alpha_2| = \alpha$, as well as the
angle between $\vec S$ and the orbital angular momentum
as a function of $\alpha$. In all fits the constant term is taken to
be zero. The data here favor a linear dependence in $\alpha$.}
\label{fig:dpdmu_v_alpha}
\end{figure}

In the following section we examine the distribution of the angle $\mu
= \hat {\vec{L}} \cdot \hat {\vec{\Delta}}$ 
that $\vec \Delta = M(\vec S_2/m_2 - \vec
S_1/m_1)$ makes with the orbital angular momentum (at $r=5M$). At
$r=50M$ this distribution is uniform (since $\vec S_1$ and $\vec S_2$
are chosen from a uniform distribution on the sphere). In
Figs.~\ref{fig:q1.00dist}-\ref{fig:q0.062dist}, we show histograms of the
distribution of the angle $\hat {\vec{L}} \cdot \hat {\vec{\Delta}}$ that $\vec \Delta$
makes with the orbital angular momentum for the given mass ratios. To
analyze these data quantitatively, we bin the data from $\mu=-1$ to
$\mu=1$ with bin widths of $\delta \mu = 0.01$. We fit the resulting
data $P(\mu)$ to a linear function $P(\mu) = P(0) +
\frac{dP}{d\mu}|_{0}
\mu$ for each mass ratio. The results are summarized in
Table~\ref{table:fits} and plots of the fits are given in
Figs.~\ref{fig:q1.00fit}-\ref{fig:q0.062fit}. We perform a similar
analysis for the angle that $\vec S$ makes with the orbital angular
momentum (see Fig.~\ref{fig:dpdmu}).  We
also perform a similar analysis, but with $q$ fixed to $q=1/4$ and
$\alpha_1=\alpha_2=\alpha$ reduced by factors of $\sqrt{2}$ and $2$,
respectively
(See Figs.~\ref{fig:q0.25smalldist}-\ref{fig:q0.25vsmalldist}
and~\ref{fig:q0.25smallfit}-\ref{fig:q0.25vsmallfit}), and fit the
resulting slope $dP/d\mu$ as a function of $\alpha$. Here the fit
favors a leading-order linear dependence in $\alpha$ over a
leading-order quadratic dependence (where the constant term is assumed
to be zero) (See Fig.~\ref{fig:dpdmu_v_alpha}).  If we set the
constant in the fit to zero, then a linear dependence is $dP /d\mu =
-(0.02491\pm0.00098) \alpha$ for the distribution of 
$\hat {\vec{L}} \cdot \hat {\vec{\Delta}}$ 
and $dP /d\mu = -(0.0301\pm0.0041)\alpha$ for the distribution
of the $\hat {\vec{L}} \cdot \hat {\vec{S}}$. 
Note that the skewing of the
distributions takes place at smaller radii, as can be seen by
differences in the 0.25S1 and 0.25F distributions, which differ only
in the orbital radius ($5M$ for 0.25S1 and $8M$ for 0.25F) where the
distributions are measured. In Table~\ref{table:afits} we show fits 
for the distributions of
the angles $\hat S_1\cdot \hat L$ and $\hat S_2\cdot \hat L$ for the
same set of runs. Note that the distribution of $\hat S_1\cdot \hat L$
(the smaller BH's spin) become essentially uniform for $q<1/4$.


\begin{table}
\caption{The distribution $P(\mu)$ of the angle $\mu=\cos \theta$
 between $\vec \Delta$ and the $\vec L$ at $r=5M$ starting from a
uniform distribution at $r=50M$ (top), and the similar distribution
for the angle between $\vec S$ and $\vec L$ (bottom). The 0.25S1
configurations had $\alpha_1 = \alpha_2 =0.97/\sqrt{2}$ and
the 0.25S2 has $\alpha = 0.97/2$, while the 0.25F configurations
have $\alpha=0.97/\sqrt{2}$ and provide the distributions at $r=8M$
(rather than $r=5M$),
 all others had
$\alpha_1 = \alpha_2 = 0.97$.}
\label{table:fits}
\begin{ruledtabular}
\begin{tabular}{ll}
$q$ & $P(\mu)$\\
\hline
1.00 & $0.5000 \pm 0.0018 +  (0.0009 \pm 0.0031) \mu$ \\
0.75 & $0.5000 \pm 0.0019 -  (0.0138 \pm 0.0034) \mu$ \\
0.50 & $0.5000 \pm 0.0019 -  (0.0180 \pm 0.0033) \mu$ \\
0.25 & $0.5000 \pm 0.0018 -  (0.0251 \pm 0.0031) \mu$ \\
0.125& $0.5000 \pm 0.0020 -  (0.0248 \pm 0.0035) \mu$ \\
0.0625 & $0.5000 \pm 0.0019 -  (0.0226 \pm 0.0033) \mu$ \\
0.25S1 & $0.5000 \pm 0.0020 - (0.0156 \pm 0.0035)\mu $\\
0.25S2 & $0.5000 \pm 0.0012 - (0.0123 \pm 0.0031)\mu $\\
0.25F & $0.5000 \pm 0.0021  - (0.0108 \pm 0.0037)\mu $\\
\hline
1.00 & $0.5000 \pm 0.0021 -  (0.0345 \pm 0.0037) \mu$ \\
0.75 & $0.5000 \pm 0.0020 -  (0.0284 \pm 0.0035) \mu$ \\
0.50 & $0.5000 \pm 0.0019 -  (0.0286 \pm 0.0031) \mu$ \\
0.25 & $0.5000 \pm 0.0019 -  (0.0261 \pm 0.0034) \mu$ \\
0.125& $0.5000 \pm 0.0018 -  (0.0249 \pm 0.0034) \mu$ \\
0.0625 & $0.5000 \pm 0.0019 -  (0.0225 \pm 0.0033) \mu$ \\
0.25S1 & $0.5000 \pm 0.0020 - (0.0162 \pm 0.0034)\mu $\\
0.25S2 & $0.5000 \pm 0.0019 - (0.0125 \pm 0.0034)\mu $\\
0.25F  & $0.5000 \pm 0.0020 - (0.0103 \pm 0.0034)\mu $\\

\end{tabular}
\end{ruledtabular}
\end{table}

\begin{table}
\caption{The distribution $P(\mu)$ of the angle $\mu=\cos \theta$
 between $\vec S_1$ and the $\vec L$ at $r=5M$ starting from a
uniform distribution at $r=50M$ (top), and the similar distribution
for the angle between $\vec S_2$ and $\vec L$ (bottom). The 0.25S1
configurations had $\alpha_1 = \alpha_2 =0.97/\sqrt{2}$ and
the 0.25S2 has $\alpha = 0.97/2$, while the 0.25F configurations
have $\alpha=0.97/\sqrt{2}$ and provide the distributions at $r=8M$
(rather than $r=5M$),
 all others had
$\alpha_1 = \alpha_2 = 0.97$. Note that the distribution of angles for
the smaller component $\vec S_1$ becomes uniform as $q\to0$.}
\label{table:afits}
\begin{ruledtabular}
\begin{tabular}{ll}
$q$ & $P(\mu)$\\
\hline
1.00 & $0.5000 \pm 0.0019 -  (0.0278 \pm 0.0033) \mu$ \\
0.75 & $0.5000 \pm 0.0020 -  (0.0129 \pm 0.0034) \mu$ \\
0.50 & $0.5000 \pm 0.0019 -  (0.0189 \pm 0.0033) \mu$ \\
0.25 & $0.5000 \pm 0.0019 -  (0.0044 \pm 0.0034) \mu$ \\
0.125& $0.5000 \pm 0.0019 -  (0.0000 \pm 0.0033) \mu$ \\
0.0625 & $0.5000 \pm 0.0019 - (0.0026 \pm 0.0033) \mu$ \\
0.25S1 & $0.5000 \pm 0.0020 - (0.0019 \pm 0.0034)\mu $\\
0.25S2 & $0.5000 \pm 0.0018 - (0.0008 \pm 0.0031)\mu $\\
0.25F & $0.5000 \pm 0.0021  - (0.0007 \pm 0.0036)\mu $\\
\hline
1.00 & $0.5000 \pm 0.0020 -  (0.0237 \pm 0.0035) \mu$ \\
0.75 & $0.5000 \pm 0.0018 -  (0.0252 \pm 0.0032) \mu$ \\
0.50 & $0.5000 \pm 0.0020 -  (0.0259 \pm 0.0034) \mu$ \\
0.25 & $0.5000 \pm 0.0020 -  (0.0261 \pm 0.0034) \mu$ \\
0.125& $0.5000 \pm 0.0019 -  (0.0249 \pm 0.0034) \mu$ \\
0.0625 & $0.5000 \pm 0.0019 -  (0.0225 \pm 0.0033) \mu$ \\
0.25S1 & $0.5000 \pm 0.0021 - (0.0162 \pm 0.0037)\mu $\\
0.25S2 & $0.5000 \pm 0.0020 - (0.0125 \pm 0.0034)\mu $\\
0.25F  & $0.5000 \pm 0.0020 - (0.0105 \pm 0.0034)\mu $\\

\end{tabular}
\end{ruledtabular}
\end{table}

An important consequence of choosing uniform distributions for the
directions of $\vec S_1$ and $\vec S_2$ (with magnitude $|\vec
S_i|=0.97$) is that the initial distributions for the squares of the
magnitudes of $\vec S$ and $\vec \Delta$, $P(S^2)$ and $P(\Delta^2)$,
are uniform in the range $[\alpha (m_2^2 - m_1^2)]^2$ to $[\alpha
(m_2^2 + m_1^2)]^2$, and zero outside this range (i.e.\ there is an
equal probability of finding any given value of $S^2$ or $\Delta^2$ in
this range).  However the distributions $P(\Delta)$ and $P(S)$
therefore contain an additional linear factor in $\Delta$ and $S$
(i.e. $P(x) = 2 x P(x^2)$ for any variable $x$), respectively. One
immediate consequence is that the distributions $P(\Delta)$ and $P(S)$
are both maximized for the largest allowed values of $S$ and $\Delta$.
Given the observation that large $\Delta$ in the orbital
plane~\cite{Campanelli:2007ew} leads to very large recoils, this bias, if
present in nature, would favor observations of large recoils. See
Sec.~\ref{sec:stats} for further analysis of the recoil distribution.

Schnittman in Ref.~\cite{Schnittman:2004vq} has studied the evolution of
spins in binary systems using orbit-averaged PN equations of motion
what allowed longer term evolutions (from separations up to $1000M$).
The results indicate strong correlations of the late angle among spins
when one starts fixing the initial direction of the spin of the primary
object and choose the secondary's spin direction at random (See Figs. 6 and 7
in \cite{Schnittman:2004vq}.) Bogdanovic et al revisit this scenario
in Ref.~\cite{Bogdanovic:2007hp} and 
find that if one is allowed to choose initial random
distributions for both spins the resulting evolution leads to close to
isotropic distributions of the late directions of the spins 
(See theirs Fig. 1). In our paper we find an small but statistically
significant bias towards counteralignment of the spins with the orbital
angular momentum (See Figs.~\ref{fig:q1.00fit}-\ref{fig:q0.25vsmallfit}.)

More recently, Herrmann {\it et al.} presented numerical studies of the PN
equations on GPUs~\cite{Herrmann:2009mr}. They  used the evolution
equation for the orbital frequency assuming quasi-circular orbits.
This equation is coupled with the spin and angular momentum precession
equations, which include the leading order spin-orbit and spin-spin
couplings.  On the other hand, in our calculation, the PN equations of
motion are derived from the Hamiltonian and include radiation
reaction effects. These have higher PN order spin-orbit and
spin-spin coupling terms.  Furthermore, the second term of the right
hand side of Eq.~(\ref{eq:RR}) has a significant effect in the PN
evolutions.  Although the evolution of $\hat{\vec L}$
in~\cite{Herrmann:2009mr} is determined only by the conservative
dynamics, we have also considered the dissipative effect due to the
radiation reaction. We find in the PN prediction that this dissipative
effect creates the statistically significant counter-alignment of the spins.

\section{Merger phase of BHBs}\label{Sec:Merger}

Unlike in the earlier inspiral phase, during the plunge and merger the
PN equations of motion do not provide a quantitatively accurate description of the
merger dynamics, and therefore do not provide robust estimates of 
the final remnant mass, spin, and recoil. However, analysis of the
recoil in particular shows that PN analysis can be used to derive
heuristic formulae (based on how PN predictions scale with spins and
masses) that give quantitatively correct
predictions~\cite{Gonzalez:2006md, Campanelli:2007ew,
Campanelli:2007cga} and incorporate the symmetries of the problem.
 We will use this modeling in the case of
the total radiated energy and angular momentum. In particular we will
supplement the inspiral losses, modeled by the energy and angular
momentum of the ISCO in the particle limit (extended to the comparable
mass regime) with the subsequent plunge using the PN dependence on
the BHs parameters (and fitting the amplitudes as in the recoil velocities
case).

\subsection{Recoil velocities}\label{SubSec:Recoil}

In order to quantify and model the nonleading corrections,
we augment our original empirical formula with
new subleading terms that are higher order in the mass ratio
and include a new term linear in the total spin, motivated by higher
order post-Newtonian computations\citep{Racine:2008kj}, and introduce 
additional parameters $B_H, B_K, H_S, K_S$ and $\Theta_1$,
\begin{eqnarray}\label{eq:Pempirical}
\vec{V}_{\rm recoil}(q,\vec\alpha)&=&v_m\,\hat{e}_1+
v_\perp(\cos\xi\,\hat{e}_1+\sin\xi\,\hat{e}_2)+v_\|\,\hat{n}_\|,\nonumber\\
v_m&=&A\frac{\eta^2(1-q)}{(1+q)}\left[1+B\,\eta\right],\nonumber\\
v_\perp&=&H\frac{\eta^2}{(1+q)}\left[
(1+B_H\,\eta)\,(\alpha_2^\|-q\alpha_1^\|)\right.\nonumber\\
&&\left.+\,H_S\,\frac{(1-q)}{(1+q)^2}\,(\alpha_2^\|+q^2\alpha_1^\|)\right],\nonumber\\
v_\|&=&K\frac{\eta^2}{(1+q)}\Bigg[
(1+B_K\,\eta)
\left|\alpha_2^\perp-q\alpha_1^\perp\right|
\nonumber \\ && \quad \times 
\cos(\Theta_\Delta-\Theta_0)\nonumber\\
&&+\,K_S\,\frac{(1-q)}{(1+q)^2}\,\left|\alpha_2^\perp+q^2\alpha_1^\perp\right|
\nonumber \\ && \quad \times 
\cos(\Theta_S-\Theta_1)\Bigg],
\end{eqnarray}
where $\eta=q/(1+q)^2$, with $q=m_1/m_2$
the mass ratio of the smaller to larger mass hole, 
$\vec{\alpha}_i=\vec{S}_i/m_i^2$, the index $\perp$ and $\|$ refer to
perpendicular and parallel to the orbital angular momentum respectively,
$\hat{e}_1,\hat{e}_2$ are
orthogonal unit vectors in the orbital plane, and $\xi$ measures the
angle between the unequal mass and spin contribution to the recoil
velocity in the orbital plane. The constants $H_S$ and $K_S$ can be determined
from newly available runs. The angle $\Theta$ is defined as the angle 
between the in-plane component of $\vec \Delta = M (\vec S_2/m_2 - \vec
S_1/m_1)$ or $\vec S=\vec S_1+\vec S_2$
and the infall direction at merger. Phases $\Theta_0$ and $\Theta_1$ depend
on the initial separation of the holes for quasicircular orbits.  

A crucial observation is that the dominant contribution to the recoil
is generated near the time of formation of the common horizon of the
merging black holes (See, for instance Fig. 6 in
~\citep{Lousto:2007db}).  The formula above (\ref{eq:Pempirical})
describing the recoil applies at this moment (more precisely, the
coefficients correspond to an averaging of the PN expressions during
the plunge phase),
 and has proven to represent
 the distribution of velocities with sufficient accuracy for astrophysical
applications.  The total recoil velocity also acquires a
correction~\citep{Sopuerta:2006et} for small eccentricities, $e$, of
the form $\vec{V}_{e}=\vec{V}_{\rm recoil}\,(1+e)$, and if one allows
for
relativistic close hyperbolic encounters, then recoils up to $10000\ \KMS$
are possible \citep{Healy:2008js}.  Although we
expect the orbits will circularize well before merger.

The most recent estimates for the above parameters can be found in
\citep{Lousto:2008dn} and references therein. The current best
estimates are: $A = 1.2\times 10^{4}\
\kms$, $B = -0.93$, $H = (6.9\pm0.5)\times 10^{3}\ \kms$,
$K=(6.0\pm0.1)\times 10^4\ \kms$, and $\xi \sim 145^\circ$.  Note that
we can use the data from~\citep{Lousto:2008dn} to obtain
$K=(6.072\pm0.065)\times 10^4\ \KMS$, if we assume that $B_{K}$ and
$K_{S}$ are negligible. 
On the other hand, by fitting the data to $K$
and $B_{K}$ simultaneously, we obtain $K=(5.24\pm0.29)\times 10^{4}\
\KMS$ and $B_{K}=0.74\pm0.29$. At first glance these two results look
quite different. However, in both cases the actual resulting
empirical
formula predict the same recoil velocities within $8\%$ over the
range
$1/10<q<1$ (the data from~\citep{Lousto:2008dn} covered the range
$1/8<q<1$).
Finally, if we fit the data to find $K$ and $K_{S}$
simultaneously we obtain $K=(6.20\pm0.12)\times 10^{4}\ \KMS$ and
$K_{S} = -0.056\pm0.041$, where we made the additional assumption that
since $\vec S=\vec \Delta$ for these runs, that $\Theta_0=\Theta_1$.
An attempt to fit all three parameters produces inaccurate fitting
parameters because the degrees of freedom in the fit and the limited
number of available runs.

Equation (\ref{eq:Pempirical}) for the recoil contains all the expected 
linear terms in the spin, and include ten fitting parameters.
Based on the works \citep{Racine:2008kj,LNZ} one could add quadratic terms,
but they are complicated expressions with more fitting parameters that
we will not include here.

\subsection{Remnant Mass}\label{SubSec:Mass}

Motivated by the success of the empirical formula for the recoil, we
propose a new empirical formula for the total radiated energy based on
the post-Newtonian equations that describe the instantaneous radiated
energy (See Eqs. (3.25) in Ref. \citep{Kidder:1995zr}, and for the
quadratic terms in the spin see Ref. \citep{Racine:2008kj}, Eq.
(5.6)). For example, the spin-spin contribution to the radiated
energy has components  quadratic in $\Delta$ that have the form,
\begin{eqnarray}\label{Expansion}
\dot{E}_{SS}&\sim& A\Delta^2+B(\hat{n}\cdot\vec{\Delta})^2
+C(\vec{v}\cdot\vec{\Delta})^2
+D(\hat{n}\cdot\vec{\Delta})(\vec{v}\cdot\vec{\Delta})\nonumber\\
&=&A(\Delta_\perp^2+\Delta_\|^2)+\Delta_\perp^2
(\tilde{B}\cos^2\theta+\tilde{C}\cos\theta\sin\theta+\tilde{D})\nonumber\\
&=&A\Delta_\|^2+b\Delta_\perp^2(\cos^2(\theta-\theta_0)+c)\nonumber\\
&=&A\Delta_\|^2+\tilde{b}\Delta_\perp^2(\cos2(\theta-\theta_0)+\tilde{c}).
\end{eqnarray}
A similar expansion
can be derived for the terms quadratic in 
$\vec{S}_0=2\vec{S}+(\delta{M}/M)\vec{\Delta}$.
In addition to the terms arising from the
instantaneous radiated energy, that allow for twelve fitting parameters,
we also included terms associated with the secular loss of energy in the
inspiral period from essentially infinite separation down to the plunge.
In order to model this contribution we adopted the form of the
the 2PN binding energy, with coefficients chosen to reproduce the 
particle limit at the ISCO\citep{Ori:2000zn}
\begin{eqnarray}
\tilde{E}_{ISCO}&=&(1-\sqrt{8}/3)+\frac{\alpha_2^\|}{18\sqrt{3}}\nonumber\\
&&-\frac{5}{324\sqrt{2}}\left[\vec\alpha^2_2-3(\alpha_2^\|)^2\right]
+{\cal O}(\alpha_2^3)\nonumber
\end{eqnarray}
where we considered terms up to quadratic order in the spin.

If we take into account the $\eta^2$ effects from self force
calculations \citep{Barack:2009ey} and 2PN effects of the 
spins (See \citep{Kidder:1995zr}, Eq.\ (4.6)), we obtain:
\begin{eqnarray}\label{EISCO}
\tilde{E}_{ISCO}&=&
(1-\sqrt{8}/3)+0.103803\eta
\nonumber\\
&&+\frac{1}{36\sqrt{3}(1+q)^2}
\left[q(1+2q)\alpha^\|_1+(2+q)\alpha^\|_2\right]
\nonumber\\
&&-\frac{5}{324\sqrt{2}(1+q)^2}\left[
\vec\alpha^2_2-3(\alpha_2^\|)^2\right.
\nonumber\\
&&\left.-2q(\vec\alpha_1\cdot\vec\alpha_2-3\alpha_1^\|\alpha_2^\|)
+q^2(\vec\alpha^2_1-3(\alpha_1^\|)^2)
\right]\nonumber\\
&&+{\cal O}(\alpha^3)
\end{eqnarray}
This expression represents a quadratic expansion in the spin-dependence,
hence we expect to produce reliable results for intrinsic spin magnitudes
$\alpha_i<0.8$. The exact expression for all values of spins, including
maximally rotating, are complicated and are given in the appendix

Thus our parametrization of the energy loss is given
\begin{eqnarray}\label{Eempirical}
&&\delta M/M = \eta\,\tilde{E}_{ISCO}
+E_2\eta^2+E_3\eta^3
\nonumber\\ &&
+\frac{\eta^2}{(1+q)^2}\Bigg\{E_S\,(\alpha_2^\|+q^2\,\alpha_1^\|)
\nonumber\\ &&
+E_\Delta\,(1-q)\,(\alpha_2^\|-q\,\alpha_1^\|)
+E_A\,|\vec\alpha_2+q\vec\alpha_1|^2
\nonumber\\ &&
+E_B\,|\alpha_2^\perp+q\alpha_1^\perp|^2
\left(\cos^2(\Theta_{+}-\Theta_2)+E_C\right) 
\nonumber\\
&&
+E_D\,|\vec\alpha_2-q\vec\alpha_1|^2
\nonumber\\ &&
+E_E\,|\alpha_2^\perp-q\alpha_1^\perp|^2
\left(\cos^2(\Theta_{-}-\Theta_3)+E_F\right)\Bigg\},
\end{eqnarray}
where $M=m_1+m_2$ and $\Theta_{\pm}$
are the angles that  $\vec\Delta_{\pm}=M (\vec S_1/m_1\pm\vec S_2/m_2)$ make 
with the radial direction
during the final plunge and merger (for comparable-mass BHs, a 
sizable fraction of the 
radiation is emitted during this final plunge, see for instance Fig. 6 in
Ref.~\citep{Lousto:2007db}). Phases $\Theta_{2,3}$ are parameters that give the
angle of maximum radiation for these terms, and depend on the initial 
separation and parameters of the binary at the beginning of the numerical
simulation. 

According to the PN theory \citep{Peters:1964}, the leading correction
to the radiated energy for small eccentricities has the form
$\dot{E}=\dot{E}_C(1+157/24e)$, where $E_C$ is the radiated energy in
the circular case, which should in principle be added to the above
formula. However, it is expected that the orbits will be
quite circularized by the time of merger.

To determine the fitting parameters in formula~(\ref{Eempirical}) we
need to correct some of the numerical data to account for energy
already lost by the system in reaching the initial separation of
the simulation. That is, some
authors choose to normalize their data such that the sum of the
horizons masses is 1, which approximates the situation where the
energy lost during the prior inspiral is taken into account, while
other normalize the initial data to unit ADM mass. In these latter
cases, we add the 3PN binding energy of the initial configuration to
the calculated radiated energy to obtain an approximation for the
total energy radiated by the system in question from infinite separation.

For the non-spinning case we fit the data found in
Refs.~\citep{Berti:2007fi, Gonzalez:2008bi}. Here we fit $E_{\rm Rad}$
versus $\eta$, where $E_{\rm Rad}$ is the total radiated energy for a
given configuration minus the binding energy of the initial
configuration (where the binding energy is negative). We calculate the
binding energy using the 3PN accurate expressions given in
\citep{Blanchet:2000nv}.
A fit of the resulting data gives  $E_2 = 0.341\pm0.014$ and $E_3 =
0.522\pm0.062$.

For the spinning cases, when spins are aligned with the orbital 
angular momentum, fits for final remnant mass from
  Ref.~\citep{Berti:2007nw, Gopakumar:2007vh} yield 
$E_{S} = 0.673\pm0.035$, $E_{\Delta} = -0.36\pm0.37$,  $E_{A} =
-0.014\pm0.021$, and $E_{D} = 0.26\pm0.44$. 
The source of these large errors is the difference in
correcting for the normalization of the results in the papers.

Finally, fits from the final remnant masses from
Ref.~\citep{Campanelli:2007cga} yields $E_E=0.09594\pm0.00045$ and
fits from the equal-mass configurations in Ref.~\citep{Lousto:2008dn}
yield $E_B = 0.045\pm0.010$. 

\subsection{Remnant Spin}\label{SubSec:Spin}

In an analogous way, we propose an empirical formula for the
final remnant spin (note that the total radiated angular momentum,
unlike the total radiated energy, is not finite for an inspiral from
infinite initial separation) 
based on the post-Newtonian equations that describe the radiated
angular momentum
(See Eqs. (3.28) in \citep{Kidder:1995zr}) and the angular
momentum of a circular binary at close separations (4.7), 
\begin{eqnarray}\label{Jempirical}
&&\vec{\alpha}_{\rm final} = 
\left(1-\delta M/M\right)^{-2}\Big\{\eta\tilde{\vec{J}}_{ISCO}
+\left(J_2\eta^2+J_3\eta^3\right)\hat{n}_\|
\nonumber\\
&& \,\,
+\frac{\eta^2}{(1+q)^2}\Big(\left[J_A\,(\alpha_2^\|+q^2\,\alpha_1^\|)
\right.\nonumber\\
&&\left. \qquad
+J_B\,(1-q)\,(\alpha_2^\|-q\,\alpha_1^\|)\right]\hat{n}_\|
\nonumber\\
&&\left. \quad 
+(1-q)\,|\vec\alpha_2^\perp-q\,\vec\alpha_1^\perp|
\right.\nonumber\\
&&\left.
\qquad \times 
\sqrt{J_\Delta\cos[2(\Theta_\Delta-\Theta_4)]+J_{M\Delta}}\,\hat{n}_\perp 
\right.\nonumber\\
&&\left. \quad
\,+|\vec\alpha_2^\perp+q^2\,\vec\alpha_1^\perp|
\right.\nonumber\\
&&
\qquad \times \sqrt{J_S\cos[2(\Theta_S-\Theta_5)]+J_{MS}}\,\hat{n}_\perp
\Big)\Big\}.
\end{eqnarray}
where we have expanded the triple cross products 
in Eq. (3.28c) in \citep{Kidder:1995zr} and used the last
form of the parametrization in Eq.~(\ref{Expansion}).

Note that, even at linear order, there are important contributions 
of generic spinning black holes producing
radiation in directions off the orbital axis that do not vanish for
equal masses nor vanishing total spin.
The above formula can be augmented by quadratic-in-the-spins
terms~\citep{Racine:2008kj,LNZ}
of a form similar to the terms added to the radiated
energy formula (\ref{Eempirical}). However, 
those terms are more
complicated and involve many more fitting constants in addition
to the ten for the linear dependence. In addition, the
linear approximations seem to have smaller quadratic corrections for
the radiated angular momentum than the radiated energy
(See for instance Fig.\ 21 of Ref.\ \citep{Campanelli:2006fy}.)

Looking at the spin expansion of the orbital angular momentum
of a particle at the ISCO \citep{Ori:2000zn},
\begin{eqnarray}
\tilde{\vec{J}}_{ISCO}&=&2\sqrt{3}\hat{n}^\|
-\frac{4}{9\sqrt{2}}\left[
\vec\alpha_2+2\alpha_2^\|\hat{n}^\|\right]\nonumber\\
&&+\frac{2}{9\sqrt{3}}
\left[\vec\alpha^2_2-3(\alpha_2^\|)^2\right]\hat{n}^\|+
\frac{1}{\eta}\frac{\vec\alpha_2}{(1+q)^2}\nonumber\\
&&+{\cal O}(\alpha_2^3).
\end{eqnarray}
and incorporating the $\eta^2$ effects from self force
calculations \citep{Barack:2009ey} and the 2PN effects of the 
spins (see Eq(4.7) in \citep{Kidder:1995zr}),
fitted to reproduce the particle limit one obtains,
\begin{eqnarray}\label{JISCO}
&&\tilde{\vec{J}}_{ISCO}=\Bigg\{2\sqrt{3}
-1.5255862\eta 
\nonumber\\
&&-\frac{1}{9\sqrt{2}(1+q)^2}
\left[q(7+8q)\alpha^\|_1+(8+7q)\alpha^\|_2\right] 
\nonumber\\
&&+\frac{2}{9\sqrt{3}(1+q)^2}\left[
\vec\alpha^2_2-3(\alpha_2^\|)^2
\right.\nonumber\\
&&\left.-2q(\vec\alpha_1\cdot\vec\alpha_2-3\alpha_1^\|\alpha_2^\|)
+q^2(\vec\alpha^2_1-3(\alpha_1^\|)^2)
\right]\bigg\}\hat{n}^\| 
\nonumber\\
&&-\frac{1}{9\sqrt{2}(1+q)^2}
\left[q(1+4q)\vec\alpha_1+(4+q)\vec\alpha_2\right] 
\nonumber\\
&&+\frac{1}{\eta}\frac{(\vec\alpha_2+q^2\vec\alpha_1)}{(1+q)^2}
+{\cal O}(\alpha^3).
\end{eqnarray}
This expression represents a quadratic expansion in the spin-dependence,
hence we expect to produce reliable results for intrinsic spin magnitudes
$\alpha_i<0.8$.
The exact expression for all values of spins, including
maximally rotating, are complicated and are given in the appendix

According to the PN theory \citep{Peters:1964}, the leading correction
to the radiated angular momentum for small eccentricities has the form
$\dot{J}=\dot{J}_C(1+23/8e)$. This correction can be added to the results,
but we expect very low eccentricities by the time the BHs merge
(see~\cite{Hinder:2007qu, Sperhake:2007gu} for the effects of
eccentricity on the remnant).

For the non-spinning case we fit the data found in
Refs.~\citep{Berti:2007fi, Gonzalez:2008bi}.
We find $J_2 = -2.81\pm0.11$ and $J_3 = 1.69\pm0.51$.
Fits for final remnant spin from  Refs.~\citep{Berti:2007nw} yield
$J_{A} = -1.971 \pm 0.018$, and $J_{B}=-3.611 \pm0.042$.  
The rest of the
fitting constants are currently hard to determine with precision.
If we attempt to fit $J_{A}$ and $J_{B}$ from the data
in~\citep{Berti:2007nw}~and~\citep{Gopakumar:2007vh}, we find
$J_{A} = -2.97\pm0.26$ and $J_{B} = -1.73\pm0.80$.
From the combined fit we find that $2.42\% < \delta M/M <
9.45\%$ and $0.34 < \alpha < 0.92$ for the equal-mass, aligned spin
scenario, in the region where the fit is valid ($|\alpha| < 0.9$).

Note that formulae  (\ref{EISCO}) and  (\ref{JISCO})
generalize the particle limit ISCO results to take into account both the mass
ratio dependence, $q=m_1/m_2$, and the spin of the holes, $S_1$ and
$S_2$,  in a symmetric way that is accurate up to quadratic
order in those parameters.  This allows us to have an accurate
description both when the binaries have relatively small mass ratios, 
i.e.\  $q\lesssim1/10$, and  in the comparable mass
regime (where the radiative terms in
(\ref{Eempirical}) and  (\ref{Jempirical}) dominate and the
ISCO is ill defined). For the full expressions see Appendix A.

\section{Merger Statistics}
\label{sec:stats}

\begin{figure}
\includegraphics[width=3.5in]{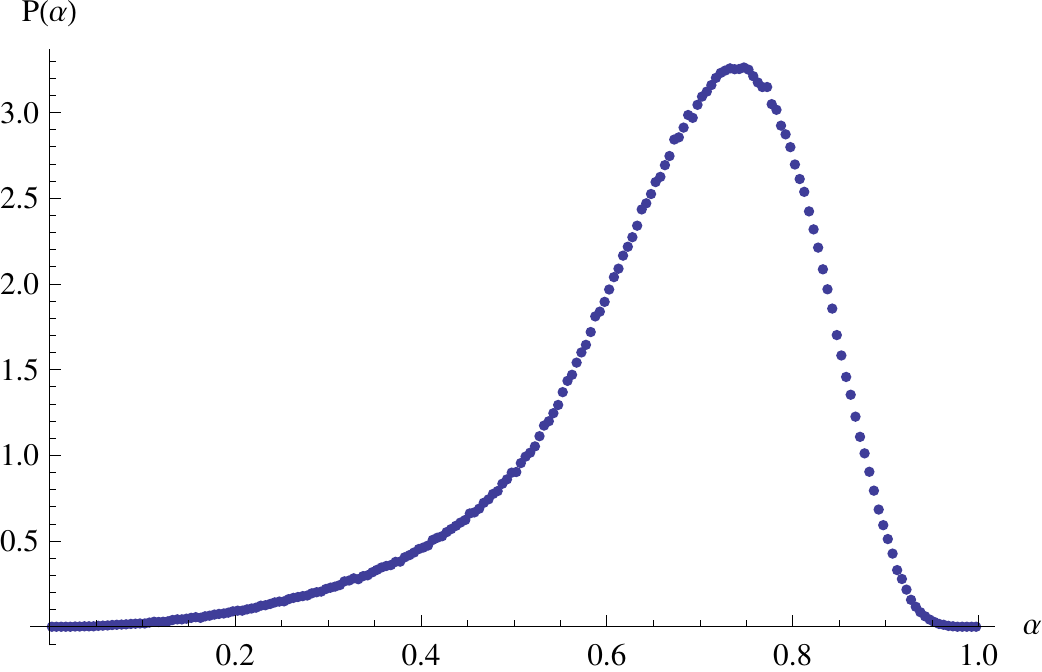}
\caption{The remnant BH spin magnitude distribution after several generation of
mergers, starting from a uniform initial distribution.}
\label{fig:sdist}
\end{figure}

In order to predict the distribution of the remnant spin and recoil, we
consider a normal random distribution of spin directions and
magnitudes and mass ratios for quasi-circular black-hole binaries
 and compute the probability
density distribution of spin magnitudes of the final remnant (see also
\citep{Berti:2008af, Tichy:2008du}).  Although our initial PN
simulations showed a slight bias towards counter-alignment of the
spins to the angular momentum, the effect is small and will be
neglected here.
The results of ten million of
such simulations are displayed in figure \ref{fig:sdist}. The final
results are insensitive to the initial distribution and quickly
converge, in a few generations of mergers, to the displayed curve,
which consequently represents a universal distribution of the intrinsic spin
magnitudes [with a maximum near 0.73 and mean in the range $(0.53,
0.83)$] of remnant BHs (when spin-up effects due accretion are 
not taken into account). We
fit the distribution of the final spins to the Kumaraswamy functional
form \citep{Jones:2009} $f(x; a,b) = a b x^{a-1}{ (1-x^a)}^{b-1}$, and
find
$a=6.58\pm0.08$, $b=7.14\pm0.19$.

\begin{figure}
\includegraphics[width=3.5in]{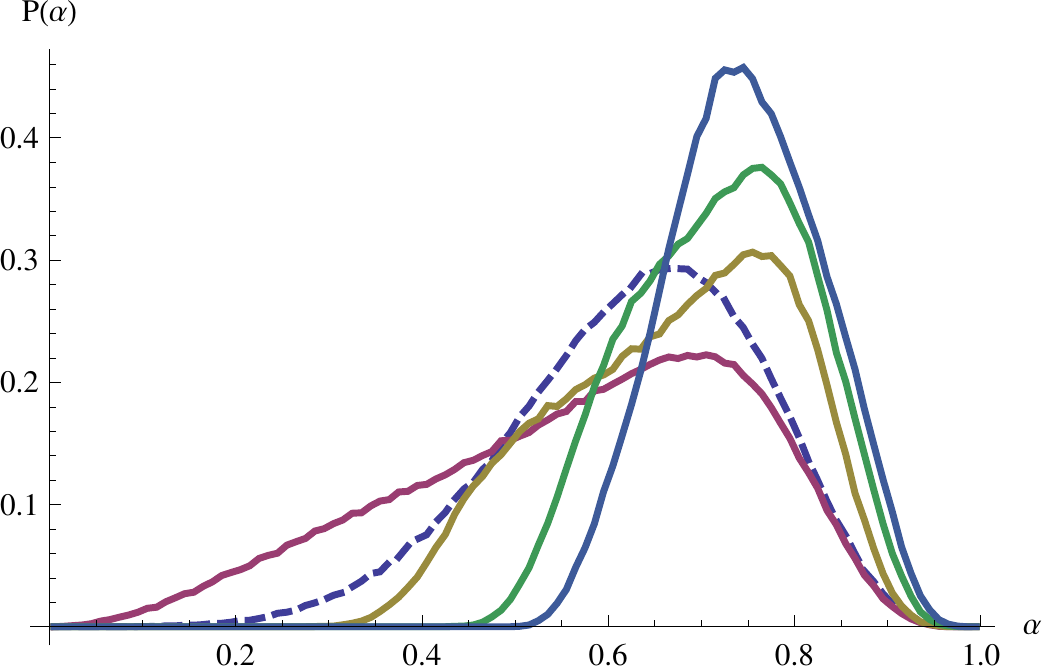}
\caption{The remnant BH spin magnitude distribution after several 
generation of mergers (starting from a uniform initial distribution) 
for mass ratios in the ranges $0\leq q\leq0.1$ (dashed), 
$0.1\leq q\leq0.2$, $0.4\leq q\leq0.5$, $0.6\leq q\leq0.7$, 
and $0.9\leq q\leq1.0$. The
distribution becomes more sharply peaked around larger values of
$\alpha$ as the mass ratio increases. Note the for the smallest range,
the spin is more highly peaked than for $0.1\leq q \leq 0.2$.}
\label{fig:qsdist}
\end{figure}

Figure \ref{fig:qsdist} show the probability distribution of the magnitude 
of the final remnant hole's spin for different ranges of the mass ratio $q$. 
We observe that the distribution becomes more and more peaked around 
the value of $\alpha\approx0.75$ as we go to comparable masses. 
The distribution gets wider for small mass ratios, this is expected 
because in the $q\to0$ limit one get essentially the original spin 
distribution of the massive black hole that was taken randomly. 
Compare this plot with a similar one in Ref.~\cite{Tichy:2008du}, 
Fig.\ 1, for dry mergers. 

\begin{figure}
\includegraphics[width=3.5in]{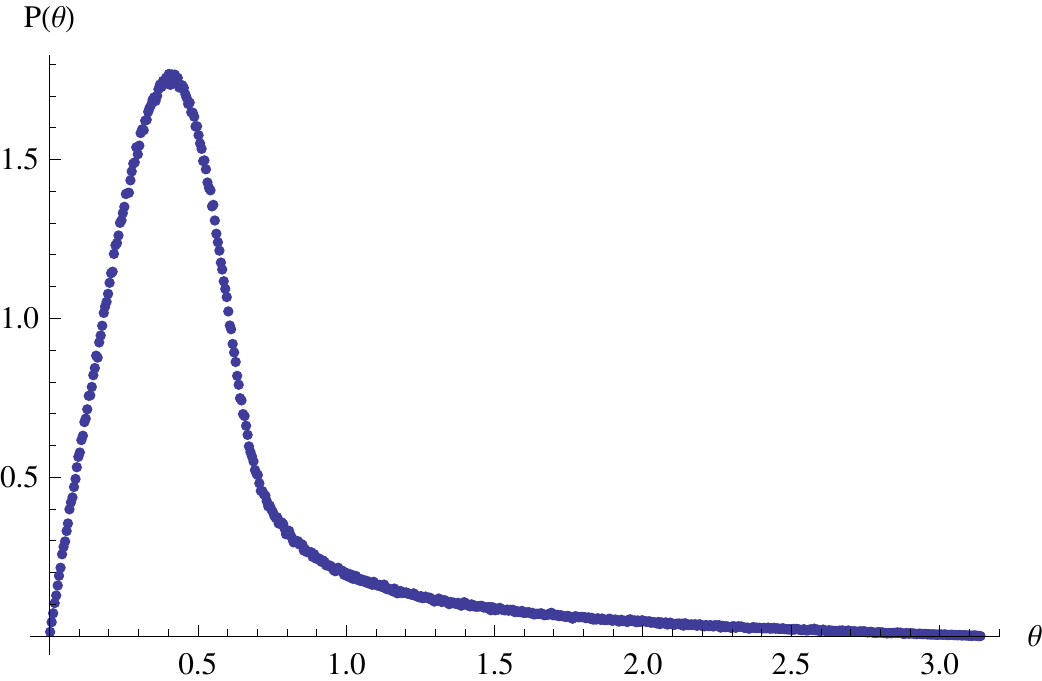}
\caption{The remnant spin direction distribution after several generation of
mergers, starting from a uniform initial distribution.}
\label{fig:stheta}
\end{figure}

\begin{figure}
\includegraphics[width=3.5in]{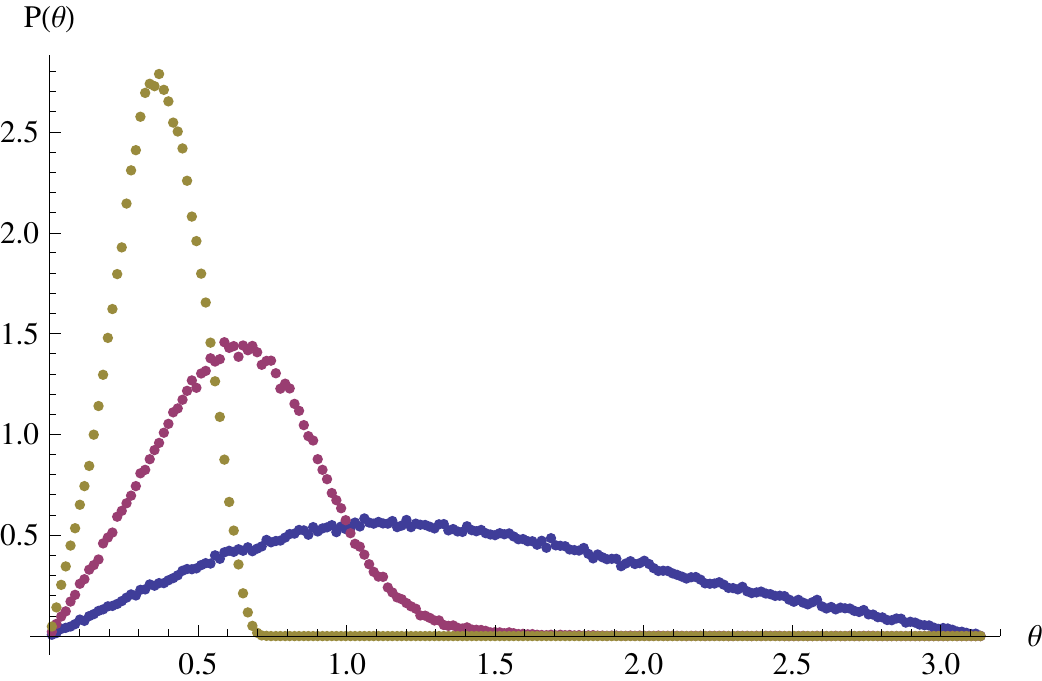}
\caption{The remnant spin direction distribution after several generation of
mergers, starting from a uniform 
initial distribution, for mass ratios in the ranges $0\leq q\leq 0.1$,
$0.2\leq q\leq0.3$, and $0.9\leq q\leq 1$. The closer to equal masses, 
the more highly peaked the distribution. For mass ratios in the range
$0.9\leq q\leq 1$, the distribution is peaked at $\theta\sim
25^\circ$. For the smaller mass ratios, the distribution approaches
$\sin \theta$ (i.e. the uniform spin direction distribution).}
\label{fig:qstheta}
\end{figure}

Figure\ \ref{fig:qstheta} displays the angular dependence 
of the probability distribution of the final spin with
respect to the original orbital angular momentum at far separations for
different ranges of mass ratios. The distribution for comparable masses is
peaked at angles close to the orbital angular momentum since the spin contributions tend to cancel leaving most of the contribution to the final angular
momentum to the orbital component. We also observe that as we go to smaller
mass ratios the distribution becomes wider since the larger black hole
contributes randomly to the final total angular momentum. We also observe
the vanishing probability of having exact alignment of the final spin with
the initial angular momentum. This is because exact alignment is a set of
measure zero on the initial random distribution of the spins.

\begin{figure}
\includegraphics[width=3.5in]{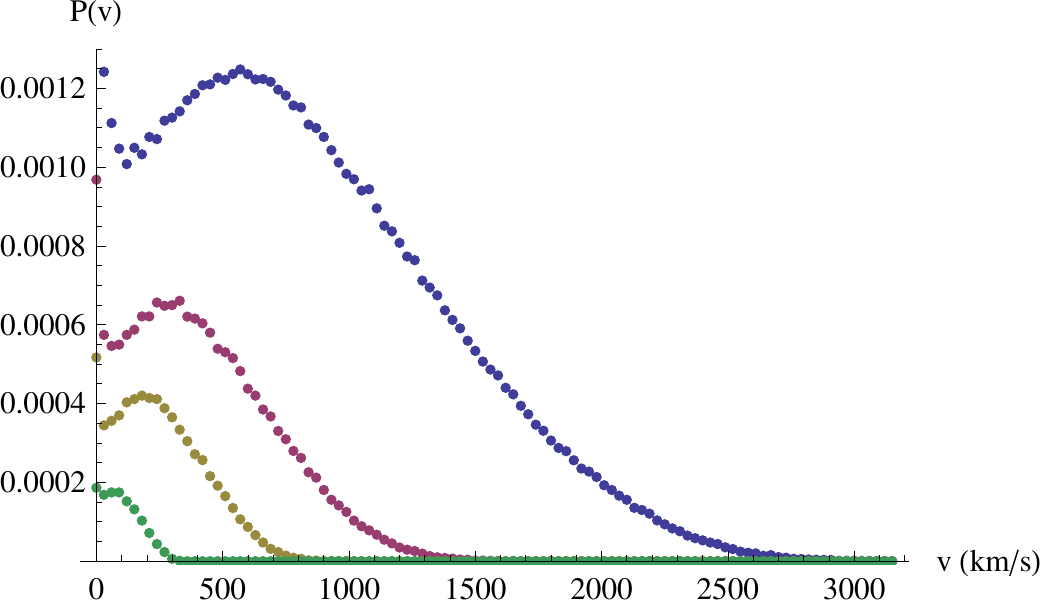}
\caption{Velocity distribution as a function of the angle the recoil
makes with the orbital angular momentum (the distribution is symmetric
with respect to $\theta \to \pi - \theta$). The plot shows $P(v)$
versus $v$ for recoils with angles in the ranges $(0,10^\circ)$ (top),
 $(10^\circ,20^\circ)$ (2nd), $(20^\circ,30^\circ)$ (3rd),
$(80^\circ,90^\circ)$ (bottom). The magnitude of $P(v)$ is the
total number of simulations with recoil-velocity directions
 between the given angles and magnitude within the range
$v\pm 15\ \KMS$.
The maxima occur at $v (\KMS)\approx 600,
300, 200, 100$, respectively. These distributions were obtained
starting from binaries with a uniform distribution in mass ratio and
distribution in spin-magnitude (random directions) given in
Fig.~\ref{fig:sdist}.}
\label{fig:v_d_theta}
\end{figure}

\begin{figure}
\includegraphics[width=3.5in]{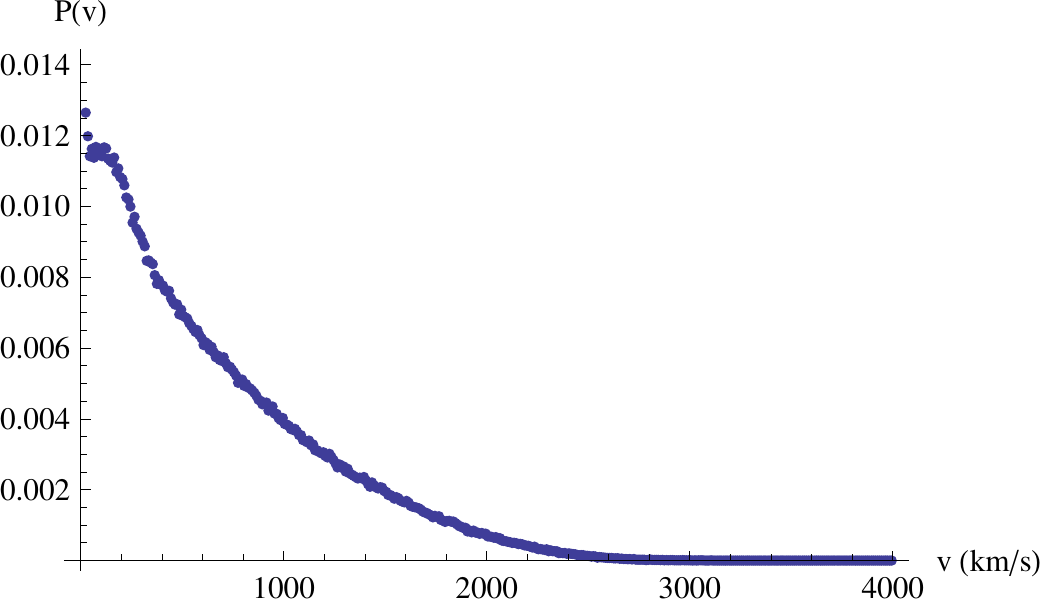}
\caption{The recoil velocity magnitude distribution for a uniform
distribution in mass ratio and spin-magnitude distribution in
Fig~\ref{fig:sdist} (with uniform spin direction). Here
$<v> = 630\ \kms$ and $\sqrt{<v^2> - <v>^2} = 534\ \kms$.}
\label{fig:vfit}
\end{figure}

We then use these distributions for the spin-magnitudes, while
assuming uniform distributions in angle and mass ratio, to predict the
distribution of recoil velocities.
In Figs.~\ref{fig:vfit}~and~\ref{fig:vtheta} we plot the distribution of
recoil magnitudes and directions, respectively. 
From the plots we see that distribution of recoils in angle is strongly
peaked toward alignment and counter-alignment with the orbital angular
momentum, which also gives the largest recoil magnitudes (this is a
consequence of the fact that the out-of-plane recoil is generally an
order magnitude larger than the in-plane recoil, as seen in
Eq.~(\ref{eq:Pempirical}) and any small component of the spins along
the orbital plane lead to those large off-plane recoils.) In
Table~\ref{table:vdist} we show probabilities for producing
various large recoils.  In Fig.~\ref{fig:v_d_theta} we
plot the distribution of recoil velocities as a function of the
angle that the recoil makes with the orbital angular momentum. Here we
find that most recoils are aligned (or counter aligned) with the
orbital angular momentum (the distributions of recoil velocities $P(v)$
at an angle $\theta$ with respect to $\hat L$ is identical to the
distribution $P(v)$ at an angle $\pi - \theta$). This strong dependence
of the magnitude of the recoil with the angle off the orbital plane
was to be expected given the large anisotropy found in the empirical
recoil formula (\ref{eq:Pempirical}) where the off-orbital plane velocities are an order
of magnitude larger than the in-plane ones (i.e. the values of the fitted
constants $H$ versus $K$.)

\begin{table}
\caption{The probability to obtain large recoil velocities, and
large recoil velocities along the line of sight.}
\label{table:vdist}
\begin{ruledtabular}
\begin{tabular}{lllll}
$v[\KMS] \geq$ & 500 & 1000 & 2000 & 2500\\
\hline
Recoil & $50.0\%$ & $22.9\%$ & $2.11\%$ & $0.21\%$\\
Observer & $22.5\%$ & $6.3\%$ & $0.21\%$ & $0.01\%$
\end{tabular}
\end{ruledtabular}
\end{table}

\begin{figure}
\includegraphics[width=3.5in]{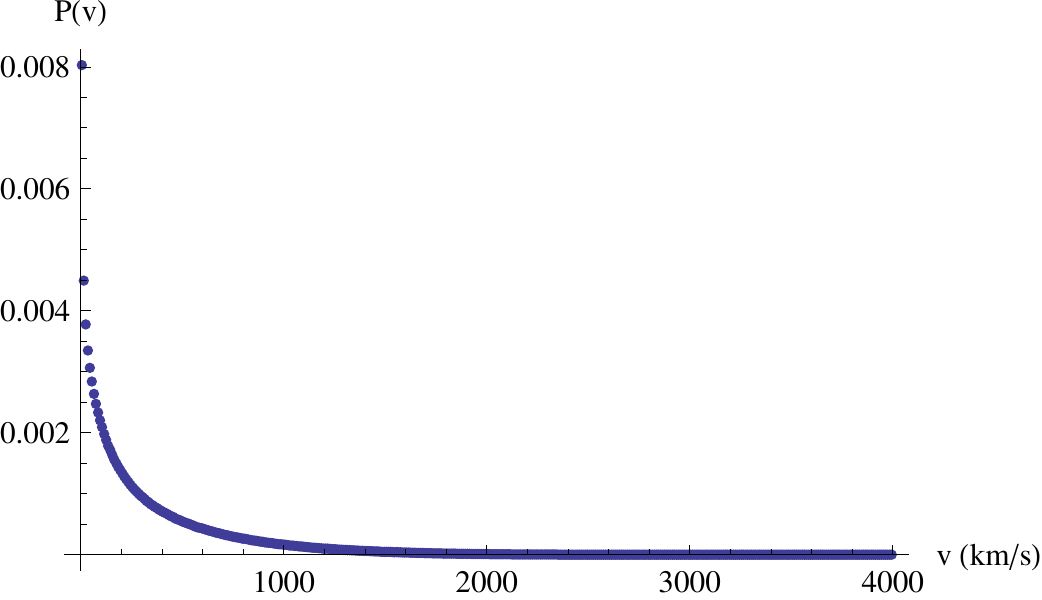}
\caption{The distribution of the recoil velocity along an observers
line of sight for a uniform
distribution in mass ratio and spin-magnitude distribution in
Fig~\ref{fig:sdist} (with uniform spin direction). Here
$<v> = 315\ \kms$ and $\sqrt{<v^2> - <v>^2} = 358\ \kms$.}
\label{fig:obs_v_dist}
\end{figure}

\begin{figure}
\includegraphics[width=3.5in]{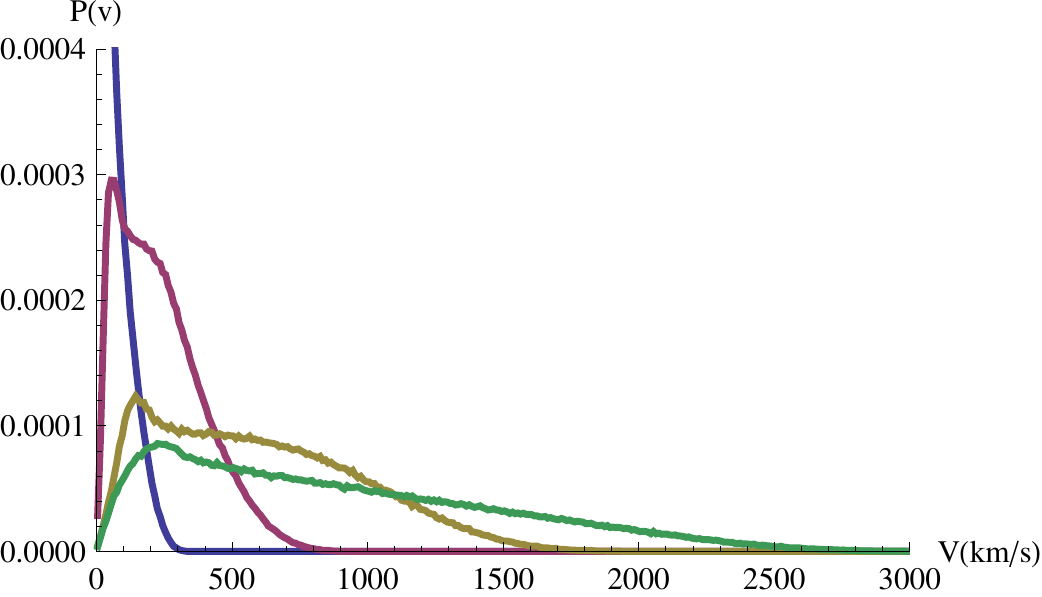}
\caption{The recoil velocity magnitude distribution for a uniform
distribution in mass ratio and spin-magnitude distribution in
Fig~\ref{fig:sdist} (with uniform spin direction). The plot shows the
recoil velocity distribution for mass ratios in the range $0\leq
q\leq 0.1$, $0.1\leq q\leq 0.2$, $0.3\leq q\leq 0.4$, and $0.9\leq q
\leq 1$. The distributions become successively broader for larger
values of $q$ (i.e.\ similar masses).}
\label{fig:qvdist}
\end{figure}

\begin{figure}
\includegraphics[width=3.5in]{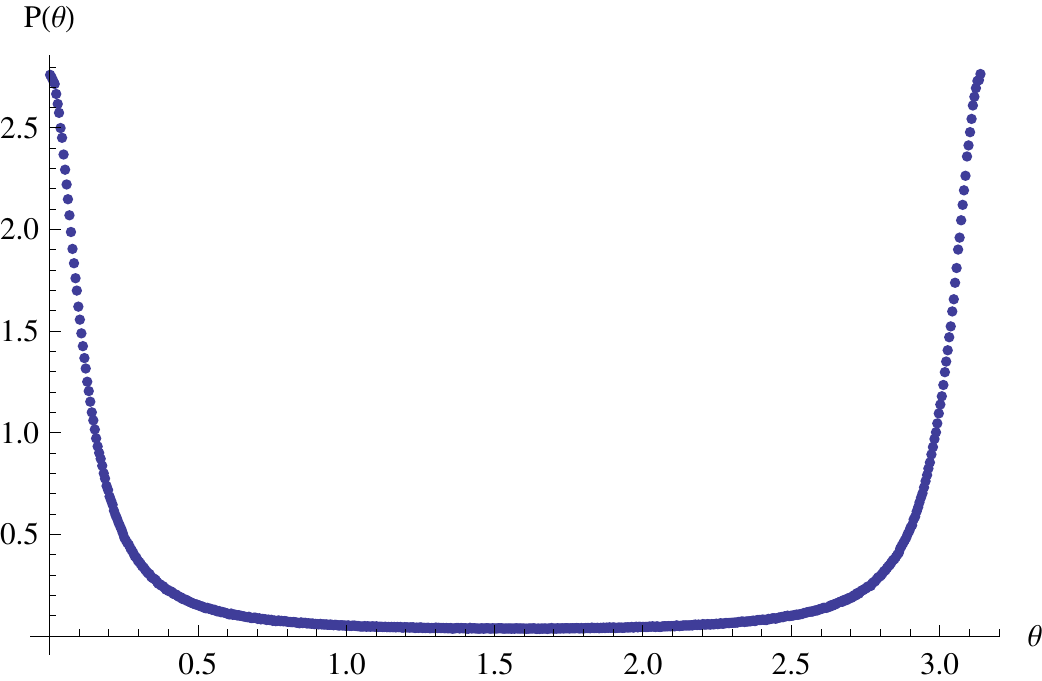}
\caption{The recoil velocity direction distribution. The angle
$\theta$ is in radians, and is defined as the angle between the recoil
vector and the orbital angular momentum vector.
}
\label{fig:vtheta}
\end{figure}

Figure\ \ref{fig:qvdist} displays the detailed dependence of the
distribution of the magnitude of the recoil velocities with the
mass ratio $q$. Here we observe that for comparable masses we obtain
a long tail of large velocity probability which recedes towards
smaller velocities as we reduce the mass ratio. This has to do
with the suppressing factor $\eta^2$ in Eqs.\ (\ref{eq:Pempirical}). 
This behavior
like $q^2$ for small mass ratios also explains why the probability
shows a peak around $v=0$. If we consider the probability density
function $P(v)$ then $dP=P(v)\,dv=P(v).(dv/dq)\,dq$. Hence
$P(v)=(dP/dq)/(dv/dq)$, and since we have chosen a white distribution
of $q$ we have $dP/dq=1$ we obtain $P(v)=1/(dv/dq)\sim1/q$, where
we have taken the leading dependence from Eq.\ (\ref{eq:Pempirical}) 
$v\sim q^2$.
This explains the sudden growth of the probability near $v=0$ when we
consider the velocity distribution in the mass ratio range $0\leq q\leq0.1$.
The wide distributions at intermediate mass ratios has to do with
the additional dependence on the direction of the spins of the holes.
This distribution drops to near zero again near the maximum recoil velocities
$v>3000\ \kms$ since this configurations require not only comparable masses,
but also counteralignment of near maximal spins that contribute with a
set of measure zero to the total probability density.

\section{Discussion}\label{sec:Discussion}

In this paper we studied the `dry' (i.e. gravitational
radiation driven) inspiral of black-hole binaries.
We performed the evolutions from separations of the order of
$50M$, using up to 3.5PN accurate expressions
 that represents an excellent approximation for this
regime (similar simulations with full numerical relativity could prove 
practically impossible with current technologies). The statistical
results show a small
bias towards counter-alignment of the vectors $\vec\Delta$ and $\vec{S}$
with respect to the orbital angular momentum $\vec{L}$ just prior to
merger. This effect
essentially takes place at close separations and
can be studied analytically at low post-Newtonian orders.
The antialignment effect is associated with the late-time precession
of the orbital plane due to radiation reaction. This effect for `dry'
mergers seems to oppose the alignment mechanism observed in `wet' mergers
\cite{Bogdanovic:2007hp,Perego:2009cw}.

After the initial inspiral regime, we studied the merger of black-hole
binaries using full numerical simulations.
Here we provided a framework to describe the bulk properties
of the remnant of a BHB merger based on PN scaling with free
parameters fixed by fitting the results  of full numerical simulations. 
We have shown how to determine
the mass loss in each BHB encounter (\ref{Eempirical}) and the spin of the
remnant BH  (\ref{Jempirical}) with fitting constants set to match
 currently available runs. Using the same
methods, one can improve the fitting parameters in the above formulae
as the results of new runs are made available,
thus providing a standard method to incorporate all 
full numerical results into astrophysical modelings.
Precessing, highly spinning binaries, including those with
small mass ratios, simulations 
can provide information to also fix the nonleading
terms in the fitting formulae. However, we expect that these extra
terms are relatively small and will not have a significant impact on
the results presented here.

The new formulae are physically motivated, as they are derived using
the post-Newtonian behavior, and naturally incorporate the correct mass ratio
$q$ dependence and physical symmetries, as well as allow
for the  radiation of angular momentum in
the orbital plane. These formulae  model
the final plunge of comparable masses black holes in an impulsive
approximation are supplemented by the slow inspiral losses that
precede this regime by adding the ISCO energy and angular momentum
in the particle limit, generalized to symmetric dependence on
the mass ratio and spins (Eqs.\ (\ref{EISCO}) and (\ref{JISCO})).

We also extended the  successful recoil formula by adding nonleading
terms to include all the linear dependence with the spins, as well
as  higher mass ratio powers in Eq.~(\ref{eq:Pempirical}).  Unlike in
the formula for the remnant recoil case, the energy and angular
momentum lost by the binary during the inspiral phase is a non-trivial
fraction of the total radiated energy and angular momentum (and, in
fact, is the dominant contribution in the small mass ratio limit). We
thus included both the instantaneous radiative terms for the plunge
phase as well as the binding
energy and angular momentum at the ISCO into our empirical formulae
(\ref{Eempirical}) and (\ref{Jempirical}).

Using the fitted coefficients in the above formulae, we find that
for equal-mass, non-spinning binaries, the net energy radiated is
$5\%$ of the total mass and the final spin is $\alpha\approx0.69$,
both in good agreement with the most accurate full
numerical runs \citep{Scheel:2008rj}. For maximally spinning BHBs
with spin aligned and counter-aligned  we estimate that quadratic corrections
lead to radiated energies between $10\%$ and $3\%$ respectively. As for
the magnitude of the remnant spin, the linear estimates are between
$0.97$ and $0.41$ respectively, with quadratic corrections slightly
reducing
those values. These results show that the cosmic censorship hypothesis
is obeyed (i.e.\ no naked singularities are formed) and are in good
agreement with earlier estimates \citep{Campanelli:2006fy}.

The set of formulae  (\ref{Eempirical}) and  (\ref{Jempirical}) with
the fitting constants determined as in the Sec.~\ref{Sec:Merger} can
be used to describe the final stage of binary black holes mergers in
theoretical, N-body, statistical studies in astrophysics and cosmology
\citep{O'Leary:2008mq, Blecha:2008mg, Miller:2008yw, Kornreich:2008ca,
Volonteri:2007dx, Gualandris:2007nm, HolleyBockelmann:2007eh,
Guedes:2008he, Volonteri:2007ax, Schnittman:2007nb, Bogdanovic:2007hp,
Schnittman:2007sn, Berti:2008af} by choosing a distribution of the
initial intrinsic physical parameters of the binaries $(q, \vec{S}_1,
\vec{S}_2)$ and mapping them to the final distribution of recoil
velocities, spins and masses after the mergers. Here we performed
initial studies and have found that: i) The merged black holes have a
considerable probability ($23\%$) to reach recoil velocities above
$1000\ \KMS$ (See Fig.~\ref{fig:vfit} and Table~\ref{table:vdist}) and
the distribution is highly peaked along the orbital angular momentum
(See Fig.~\ref{fig:vtheta}). ii) The direction of the spin of the
final merged black holes is strongly peaked at an angle of
$\approx25^\circ$ with respect to the orbital angular momentum pre-merger (see
Fig.~\ref{fig:stheta}), and the spin magnitude is strongly peaked 
at $S_f/M_f^2\approx0.73$ (see Fig.~\ref{fig:sdist}). Higher spins are
likely if we include the effects of accretion. This information can
be useful in modeling the observational effects of supermassive black
holes kicked out of their host galaxies~\cite{Bogdanovic:2009xd}.
For example, as a first approximation, one may assume that 
the inner
accretion disk is associated with the orbital plane 
of the merging binary, while the direction of the final spin is
associated  with the
current direction of the radio-jet, and finally that the preferred direction
of the kick is along the orbital angular momentum. We can then
to reconstruct 3D recoil velocities out of the observer (redshift
velocities) information.
Also, when modeling the effects of kicks on accretion disks surroundings
the merged binary, one should take into account that the most likely
recoil velocity depends on the angle with respect to the binary's orbital
plane (See Fig. \ref{fig:v_d_theta}).

In order to take into account that one of the main methods to search for
recoiling black holes is to look for large differential redshift, typically
between narrow band emission lines coming from the host galaxy and broad
emission lines from the portion of the accretion disk that the recoiling
black hole carries with it 
\cite{Komossa:2008qd, Strateva:2008wt,Shields:2009jf}, 
we have computed the effect of projecting
the computed recoil velocity from the merger of two black holes along the
line of sight of an observer on earth. The results are plotted in
Fig. \ref{fig:obs_v_dist} and in Table \ref{table:vdist}.

\begin{figure}
  \caption{The distribution of recoil velocity magnitudes and
directions (with respect to the orbital plane) assuming a distribution
in mass ratios uniform in $\log_{10} q$ in the interval
$-2\leq\log_{10}q\leq0$.}
  \includegraphics[width=3.5in]{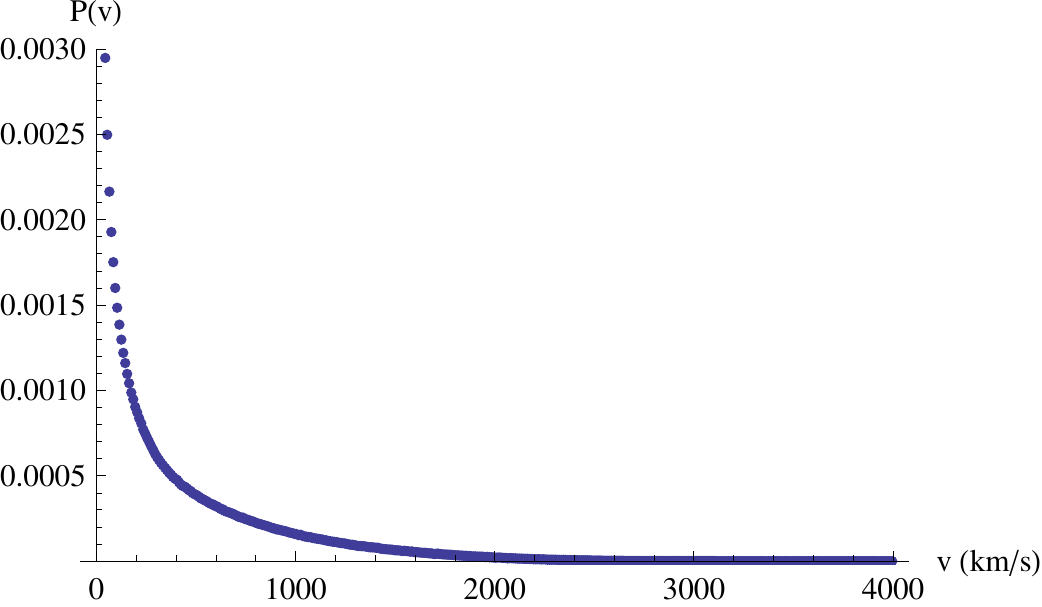}
  \includegraphics[width=3.5in]{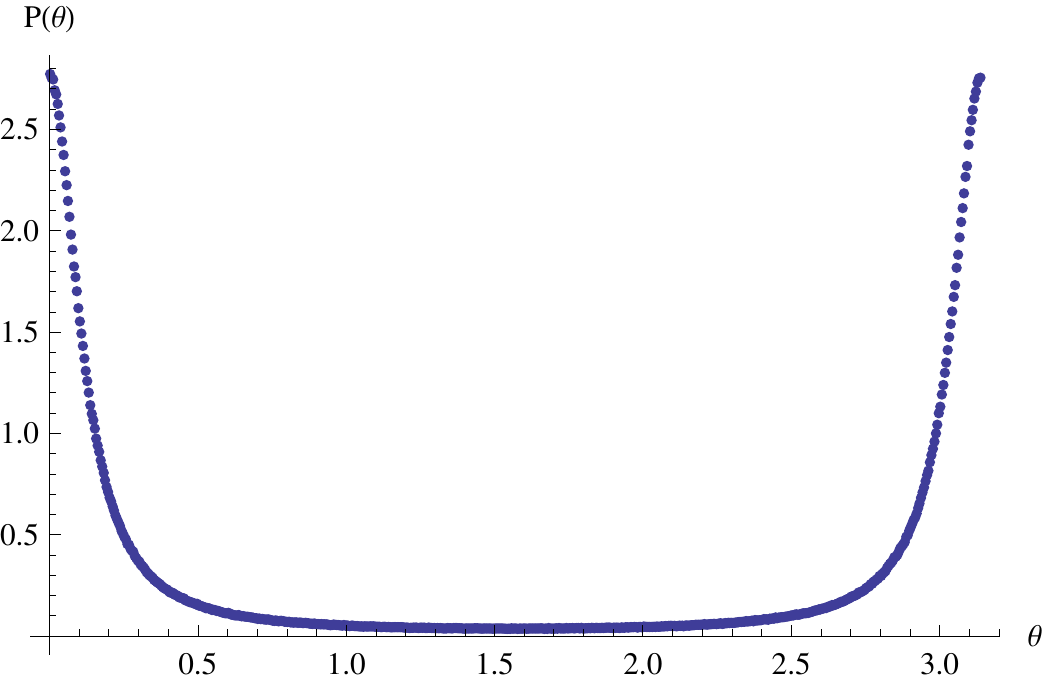}
\label{fig:logqdistkick}
\end{figure}

Finally, we note that the recoil distributions are sensitive to the 
assumed distribution of
mass ratios. Ideally one should use a distribution consistent
with the true distribution of mass ratios of merging galaxies. 
In~\cite{Volonteri:2008gj} this distribution is derived analytically
from merger scenarios in a cosmological model. They find that the
distribution is nearly flat in $\log_{10} q$ from $q=1/100$ to $q=1$
(see Ref.~\cite{Volonteri:2008gj}, Fig 1.).  
When we use a
distribution uniform in $\log_{10} q$ from $-2\leq \log_{10}q\leq0$
here,
as suggested in~\cite{Volonteri:2008gj}, the expected recoil velocity
distribution is skewed towards much lower velocities. In
Fig.~\ref{fig:logqdistkick} we plot the recoil magnitude distribution
for this distribution, while in Table~\ref{table:logqvdist} 
we give the probabilities for obtaining large recoil velocities with
this mass ratio distribution. Note that this distribution is, in fact, 
strongly 
skewed towards lower mass ratios when compared to the distribution
uniform in $q$. Consequently we see much lower probabilities for large
recoils.
\begin{table}
\caption{The probability to obtain large recoil velocities, and
large recoil velocities along the line of sight if the distribution in
mass ratios is uniform in $\log_{10} q$ for $-2\leq\log_{10}q\leq 0$.}
\label{table:logqvdist}
\begin{ruledtabular}
\begin{tabular}{lllll}
$v[\KMS] \geq$ & 500 & 1000 & 2000 & 2500\\
\hline
Recoil & $20.95\%$ & $7.99\%$ & $0.60\%$ & $0.06\%$\\
Observer & $8.62\%$ & $2.05\%$ & $0.06\%$ & $0.003\%$
\end{tabular}
\end{ruledtabular}
\end{table}

\acknowledgments 

We thank E.Bonning, A.Robinson, J. Schnittman, and M.Volonteri 
for interesting discussions.
We gratefully acknowledge NSF for financial support from grant
PHY-0722315, PHY-0653303, PHY-0714388, PHY-0722703, DMS-0820923, and
PHY-0929114; and NASA for financial support from grant NASA
07-ATFP07-0158 and HST-AR-11763.  Computational resources were
provided by Ranger cluster at TACC (Teragrid allocations TG-PHY080040N
and TG-PHY060027N) and by NewHorizons at RIT.

\appendix

\section{Innermost stable circular orbit of "Kerr" geodesics}

In this appendix we provide the necessary formulae to compute
the $E_{ISCO}$ and $J_{ISCO}$ denoted in the empirical equations
for the black hole remnant mass and spin, i.e. Eqs.\ (\ref{Eempirical}) and
(\ref{Jempirical}). Note that in the main text we have given explicitly
those functions up to quadratic terms in the spin.
We also provide explicit analytic expressions of the ISCO radius
for equatorial and polar orbits (See Eq.\ (\ref{eq:rISCOPolar}).)

When we treat the spin effects on the equations of motion 
in the effective one body (EOB) approach, 
it is useful to define two combinations of the spins~\cite{Damour:2001tu}, 
$\vec{S}_0$ in Eq.~(\ref{eq:Def_PN}) and $\vec{\tilde S}$, 
\begin{eqnarray}
\vec{\tilde S} &=& \vec{S} + {\frac{\delta M}{M}} \vec{\Delta}
\nonumber \\ 
&=& \frac{m_2}{m_1}\vec{S}_1 + \frac{m_1}{m_2}\vec{S}_2 \,,
\end{eqnarray} 
This is because we can rewrite $\vec{\tilde S}$ by using 
the nondimensional spins, $\vec{\alpha}_1$ and $\vec{\alpha}_2$,
\begin{eqnarray}
\vec{\tilde S} 
&=& {m_1}{m_2}\left( \vec{\alpha}_1 + \vec{\alpha}_2 \right) \,,
\end{eqnarray}
and then when we consider $\eta=0$, $m_1$ or $m_2$ is zero. 
We note that this definition of $\vec{\tilde S}$ is same as 
$\boldsymbol{\sigma}$ in~\cite{Damour:2001tu} 
except the numerical coefficient. 

Based on the equations given in~\cite{Sago:2005fn}, 
we focus only on $\vec{S}_0$ to derive  
the innermost stable circular orbit (ISCO) 
of the Kerr spacetime with a spin $a=S_0/M$. 
Here, we assume that the direction of $\vec{S}_0$ along the $z$-axis 
and we use the Boyer-Lindquist coordinates.  
In practice, we should use and write "spherical" orbits 
due to the spins of a binary. But here, we call "circular" orbits. 

When we consider the geodesic motion, 
$z^{\alpha}(\tau) = \{t_z(\tau),r_z(\tau),\theta_z(\tau),\phi_z(\tau)\}$, 
where $\tau$ is the proper time along the orbit,  
there are three constants of motion as follows.  
\begin{eqnarray}
E &=& -u^{\alpha}\xi_{\alpha}^{(t)} 
\,, \nonumber \\
L_z &=& u^{\alpha}\xi_{\alpha}^{(\phi)} 
\,, \nonumber \\
Q &=& K_{\alpha\beta}u^{\alpha}u^{\beta}
\,, 
\end{eqnarray}
where $u^{\alpha} = dz^{\alpha}/d\tau$, 
and the two Killing vectors are 
$\xi_{(t)}^{\mu} = (\partial_t)^\mu$ and  
$\xi_{(\varphi)}^{\mu} = (\partial_\varphi)^\mu$.
And also, we define the Killing tensor, 
$K_{\mu\nu} = 2\Sigma \,l_{(\mu}n_{\nu)}+r^2 g_{\mu\nu}$,  
where 
$l^{\mu} = \left({r^2+a^2},\Delta,0,a \right)/\Delta$ 
and 
$n^{\mu} = \left(r^2+a^2,-\Delta,0,a\right)/(2\Sigma)$
are two radial null vectors. 
Here, $\Delta = r^2-2Mr+a^2$ and $\Sigma = r^2+a^2\cos^2\theta$. 
The Killing tensor satisfies the equation $K_{(\mu\nu;\rho)}=0$. 
We also define another notation for the Carter constant, 
$C = Q-(aE-L_z)^2$. 
For equatorial plane orbits, $C$ defined by this vanishes. 

The energy per unit mass for circular orbit with the radius $r_{\rm BL}$ is derived as 
\begin{widetext}
\begin{eqnarray}
E &=& 
\biggl[
2\, Ma \left( {r_{\rm BL}}^{2}-2\,Mr_{\rm BL}+{a}^{2} \right) 
 \left( -{r_{\rm BL}}^{2}y-{r_{\rm BL}}^{2}+{a}^{2}y \right) 
\nonumber \\ && \times 
\sqrt {M \left( {a}^{4}y+2\,{r_{\rm BL}}^{2}y{a}^{2}-4\,Mr_{\rm BL}\,y{a}^{2}
+{r_{\rm BL}}^{4}+{r_{\rm BL}}^{4}y \right) {r_{\rm BL}}^{3}}
\nonumber \\ && 
/
\left( 
 \left( 2\,y{a}^{4}Mr_{\rm BL}+y{a}^{4}{r_{\rm BL}}^{2}+y{a}^{4}{M}^{2}
+2\,y{a}^{2}{r_{\rm BL}}^{4}-6\,y{r_{\rm BL}}^{2}{a}^{2}{M}^{2}
-4\,y{a}^{2}M{r_{\rm BL}}^{3}-6\,{r_{\rm BL}}^{5}My+{r_{\rm BL}}^{6}y
\right. \right. 
\nonumber \\ && \quad 
\left. \left. 
+9\,{M}^{2}{r_{\rm BL}}^{4}y-4\,{a}^{2}M{r_{\rm BL}}^{3}
-6\,{r_{\rm BL}}^{5}M
+{r_{\rm BL}}^{6}+9\,{M}^{2}{r_{\rm BL}}^{4} \right) 
\left( {a}^{4}y+2\,{r_{\rm BL}}^{2}y{a}^{2}+{r_{\rm BL}}^{4}
+{r_{\rm BL}}^{4}y \right) \right)
\nonumber \\ && 
+ 
r_{\rm BL}\, \left( 2\,{r_{\rm BL}}^{9}y
+{r_{\rm BL}}^{9}{y}^{2}+{r_{\rm BL}}^{9}+2\,{r_{\rm BL}}^{5}{a}^{4}y
+6\,{r_{\rm BL}}^{5}{y}^{2}{a}^{4}+4\,{r_{\rm BL}}^{7}y{a}^{2}
+4\,{r_{\rm BL}}^{7}{y}^{2}{a}^{2}+r_{\rm BL}\,{a}^{8}{y}^{2}
\right. 
\nonumber \\ && \quad 
\left. 
-23\,{r_{\rm BL}}^{6}My{a}^{2}-12\,{r_{\rm BL}}^{4}M{a}^{4}y
+33\,{M}^{2}{r_{\rm BL}}^{5}y{a}^{2} 
+14\,{M}^{2}{r_{\rm BL}}^{3}{a}^{4}y
-20\,{r_{\rm BL}}^{6}{y}^{2}M{a}^{2}-18\,{r_{\rm BL}}^{4}{y}^{2}M{a}^{4}
\right. 
\nonumber \\ && \quad \left. 
+28\,{M}^{2}{r_{\rm BL}}^{5}{y}^{2}{a}^{2}
+8\,{M}^{2}{r_{\rm BL}}^{3}{y}^{2}{a}^{4}
-4\,{r_{\rm BL}}^{2}{y}^{2}{a}^{6}M-8\,{M}^{3}{r_{\rm BL}}^{4}y{a}^{2}
-4\,{M}^{3}{r_{\rm BL}}^{2}{a}^{4}y
-3\,{a}^{6}M{r_{\rm BL}}^{2}y
\right. 
\nonumber \\ && \quad \left. 
+{a}^{6}{M}^{2}r_{\rm BL}\,y-8\,{M}^{3}{r_{\rm BL}}^{4}{y}^{2}{a}^{2}
+4\,{M}^{3}{r_{\rm BL}}^{2}{y}^{2}{a}^{4}
-4\,{M}^{2}r_{\rm BL}\,{y}^{2}{a}^{6}
-3\,{r_{\rm BL}}^{6}{a}^{2}M-14\,{r_{\rm BL}}^{8}My
\right. 
\nonumber \\ && \quad \left. 
+5\,{M}^{2}{r_{\rm BL}}^{5}{a}^{2}
+32\,{M}^{2}{r_{\rm BL}}^{7}y-7\,{r_{\rm BL}}^{8}{y}^{2}M
+16\,{M}^{2}{r_{\rm BL}}^{7}{y}^{2}+4\,{r_{\rm BL}}^{3}{y}^{2}{a}^{6}
-24\,{M}^{3}{r_{\rm BL}}^{6}y
-12\,{M}^{3}{r_{\rm BL}}^{6}{y}^{2}
\right. 
\nonumber \\ && \quad \left. 
+{a}^{8}{y}^{2}M 
-7\,{r_{\rm BL}}^{8}M+16\,{M}^{2}{r_{\rm BL}}^{7}-12\,{M}^{3}{r_{\rm BL}}^{6} \right) 
/
\left(
 \left( 2\,y{a}^{4}Mr_{\rm BL}+y{a}^{4}{r_{\rm BL}}^{2}+y{a}^{4}{M}^{2}
+2\,y{a}^{2}{r_{\rm BL}}^{4}
\right. \right. 
\nonumber \\ && \quad \left. \left. 
-6\,y{r_{\rm BL}}^{2}{a}^{2}{M}^{2}
-4\,y{a}^{2}M{r_{\rm BL}}^{3}-6\,{r_{\rm BL}}^{5}My
+{r_{\rm BL}}^{6}y
+9\,{M}^{2}{r_{\rm BL}}^{4}y-4\,{a}^{2}M{r_{\rm BL}}^{3}
-6\,{r_{\rm BL}}^{5}M
\right. \right. 
\nonumber \\ && \quad \left. \left. 
+{r_{\rm BL}}^{6}
+9\,{M}^{2}{r_{\rm BL}}^{4} \right)  \left( {a}^{4}y
+2\,{r_{\rm BL}}^{2}y{a}^{2}+{r_{\rm BL}}^{4}+{r_{\rm BL}}^{4}y \right) 
\right)
\biggr]^{1/2} \,.
\label{eq:fullE}
\end{eqnarray}
\end{widetext}
Here, $y$ is introduced as a dimensionless inclination parameter 
defined by 
\begin{eqnarray}
y =  \frac{C}{L_z^2} \,.
\label{eq:ydef}
\end{eqnarray}
This is related to an inclination angle as 
\begin{eqnarray}
\cos \,\iota_{\rm BL} &=& \frac{1}{\sqrt{y+1}} 
\nonumber \\ 
&=& \frac{L_z}{\sqrt{C+L_z^2}} \,,
\label{eq:Ydef}
\end{eqnarray}
and the inclination angle gives the exact inclination in 
the case of the Newtonian orbit. 

It should be noted that although we can define the circular orbit 
as a orbit with a constant radius in the Boyer-Lindquist coordinates, 
the circular orbit in another coordinates has a time dependent radius~\cite{Racine:2008kj}. 
The detail analysis of the gauge transformation 
have been discussed in~\cite{Hergt:2007ha}. 

And then, the angular momentum per unit mass along the z-axis is calculated by 
\begin{eqnarray}
L_z^2 &=& 
{\frac {{r_{\rm BL}}^{2} \left( {a}^{2}+3\,{r_{\rm BL}}^{2} \right) {E}^{2}}
{{r_{\rm BL}}^{2}y+{r_{\rm BL}}^{2}-{a}^{2}y}}
\nonumber \\ && 
-{\frac {{r_{\rm BL}}^{2} 
\left( 3\,{r_{\rm BL}}^{2}-4\,Mr_{\rm BL}+{a}^{2} \right) }{{r_{\rm BL}}^{2}y
+{r_{\rm BL}}^{2}-{a}^{2}y}} 
\,.
\label{eq:fullL}
\end{eqnarray}
Here and hereafter, we focus only on the case that $L_z \geq 0$. 

The ISCO radius in the Kerr spacetime is obtained by solving the following equation 
with respect to $r_0$. 
\begin{widetext}
\begin{eqnarray}
0 &=& 
\left( -6\,{r_0}^{5}M+4\,{a}^{2}M{r_0}^{3}-6\,{a}^{4}Mr_0
+{a}^{6}+3\,{a}^{2}{r_0}^{4}+3\,{a}^{4}{r_0}^{2}
+{r_0}^{6} \right) ^{2}{y}^{2}
\nonumber \\ 
&& -2\,{r_0}^{4} \left( 3\,{a}^{8}
-12\,r_0\,{a}^{6}M+8\,{r_0}^{2}{a}^{6}+28\,{a}^{4}{M}^{2}{r_0}^{2}
-60\,{r_0}^{3}M{a}^{4}+6\,{a}^{4}{r_0}^{4}+24\,{M}^{2}
{r_0}^{4}{a}^{2} \right. 
\nonumber \\ 
&& \left. \quad 
+28\,{r_0}^{5}{a}^{2}M-36\,{M}^{2}{r_0}^{6}
+12\,{r_0}^{7}M-{r_0}^{8} \right) y
\nonumber \\ 
&& +{r_0}^{8} 
\left( 9\,{a}^{4}-28\,{a}^{2}Mr_0-6\,{r_0}^{2}{a}^{2}
+36\,{M}^{2}{r_0}^{2}-12\,M{r_0}^{3}+{r_0}^{4} \right) 
\,.
\label{eq:ISCOFINDER}
\end{eqnarray}
\end{widetext}

In the following, we discuss the equatorial and polar orbits analytically. 
In general inclined orbit cases, we need to solve Eq.~(\ref{eq:ISCOFINDER}) numerically. 

\subsection{Equatorial circular orbit}

For the equatorial orbit, we may consider $y=0$ 
in Eq.~(\ref{eq:ISCOFINDER}). 
\begin{eqnarray}
0 &=& ( 9\,{a}^{4}-28\,{a}^{2}Mr_0-6\,{r_0}^{2}{a}^{2}
+36\,{M}^{2}{r_0}^{2}
\nonumber \\ && 
-12\,M{r_0}^{3}+{r_0}^{4} ) 
\nonumber \\ 
&=& \left( 3\,{a}^{2}-{r_0}^{2}+8\,\sqrt {M}\sqrt {r_0}a+6\,Mr_0 \right) 
\nonumber \\ && \times 
\left( 3\,{a}^{2}-{r_0}^{2}-8\,\sqrt {M}\sqrt {r_0}a+6\,Mr_0 \right) 
\,.
\end{eqnarray}
This case has been discussed analytically in~\cite{Bardeen:1972fi} 
which we summarize below. 
It should be noted that we consider the parameter range $-M \leq a \leq M$ 
and only the orbits with $L_z>0$ in our treatment. 

The appropriate ISCO solution is derived from 
\begin{eqnarray}
0 &=& {r_0}^{2}-6\,Mr_0+8\,M^{3/2}\chi\,\sqrt {r_0}-3\,{M}^{2}\chi^2
\,,
\end{eqnarray}
where $\chi=a/M$. The solution of the above quartic function is obtained 
as follows. For $\chi>0$, we have
\begin{eqnarray}
&& r_{ISCO} = M 
\left\{3+\sqrt {3 \chi^2+{\lambda}^{2}}
\right. \nonumber \\ && \left. 
-\left[(3-\lambda)(3+\lambda+2\,\sqrt {3 \chi^2+{\lambda}^{2}})\right]^{1/2} \right\} \,,
\label{eq:ISCOp}
\end{eqnarray}
and for $\chi<0$,
\begin{eqnarray}
&& r_{ISCO} = M 
\left\{3+\sqrt {3 \chi^2+{\lambda}^{2}}
\right. \nonumber \\ && \left. 
+\left[(3-\lambda)(3+\lambda+2\,\sqrt {3 \chi^2+{\lambda}^{2}})\right]^{1/2} \right\} \,,
\label{eq:ISCOm}
\end{eqnarray}
where
\begin{eqnarray}
\lambda &=& 1+(1-{\chi}^{2})^{1/3} 
\nonumber \\ && \times 
\left[ (1+\chi)^{1/3}+(1-\chi)^{1/3} \right] \,.
\end{eqnarray}
We can obtain the well known result, 
$r_{ISCO}=M$ for $a=M$ and $r_{ISCO}=9M$ for 
$a=-M$ from Eqs.~(\ref{eq:ISCOp}) and (\ref{eq:ISCOm}), respectively.

\subsection{Polar circular orbit}

For the polar orbit, we need to consider the limit $y \to \infty$ 
in Eq.~(\ref{eq:ISCOFINDER}). This means that we may solve 
the following equation. 
\begin{eqnarray}
0 &=& 
\left({r_0}^{6} -6\,{r_0}^{5}M+3\,{a}^{2}{r_0}^{4}
+4\,{a}^{2}M{r_0}^{3}
\right. \nonumber \\ && \left. \quad 
+3\,{a}^{4}{r_0}^{2}-6\,{a}^{4}Mr_0+{a}^{6}
 \right) \,.
\end{eqnarray}

Here, we introduce two nondimensional variables, $\tilde r$ and $\chi$ as 
\begin{eqnarray}
r_0 &=& M \chi \tilde r \,, 
\nonumber \\
a &=& M \chi \,.
\end{eqnarray}
Here we consider the case for $\chi \neq 0$. 
Then, the equation to find the ISCO radius is written as
\begin{eqnarray}
0 &=& {\tilde r}^{6}-6\,{\frac {{\tilde r}^{5}}{\chi}}+3\,{\tilde r}^{4}
+4\,{\frac {{\tilde r}^{3}}{\chi}}
\nonumber \\ && 
+3\,{\tilde r}^{2}
-6\,{\frac {\tilde r}{\chi}}+1 \,.
\end{eqnarray}
The solutions are nice relations (we can find them from numerical method), 
and the solutions are given by 
\begin{eqnarray}
&& r_s\,,\, \frac{1}{r_s}\,,\, \exp(i \theta_1)\,,\, \exp(-i \theta_1)\,,
\nonumber \\ && 
\exp(i \theta_2)\,,\, \exp(-i \theta_2)\,,
\end{eqnarray}
where $r_s$, $\theta_1$ and $\theta_2$ are real. 
The above solutions suggest that the 6th order equation 
can be reduced to 
\begin{eqnarray}
0 &=& ({\tilde r}^2-\alpha_1 {\tilde r}+1) 
({\tilde r}^2-\alpha_2 {\tilde r}+1) 
\nonumber \\ && \times 
({\tilde r}^2-\alpha_3 {\tilde r}+1) \,.
\end{eqnarray}
Here, although $\alpha_1$ is real, $\alpha_2$ and $\alpha_3$ are 
complex and complex conjugate each other. Therefore, we focus on 
the equation with the real coefficient, 
\begin{eqnarray}
0 &=& ({\tilde r}^2-\alpha_1 {\tilde r}+1) \,,
\end{eqnarray}
where we find 
\begin{eqnarray}
\alpha_1 &=& 2\,
{\frac {\left(1-{\chi}^{2}+i\chi\sqrt {2-{\chi}^{2}}\right)^{1/3}}{\chi}}
\nonumber \\ && 
+2\,{\frac {1}{\chi\,\left(1-{\chi}^{2}+i\chi\sqrt {2-{\chi}^{2}}\right)^{1/3}}}
+2\,\frac{1}{\chi}
\nonumber \\ 
&=& \frac{2}{\chi}
\left(\exp(i \beta/3) + \exp(- i \beta/3) +1 \right) \,;
\nonumber \\ 
\tan \beta &=& \frac{\chi\sqrt {2-{\chi}^{2}}}{1-{\chi}^{2}} \,.
\end{eqnarray}

\begin{figure}
\includegraphics[width=3.5in]{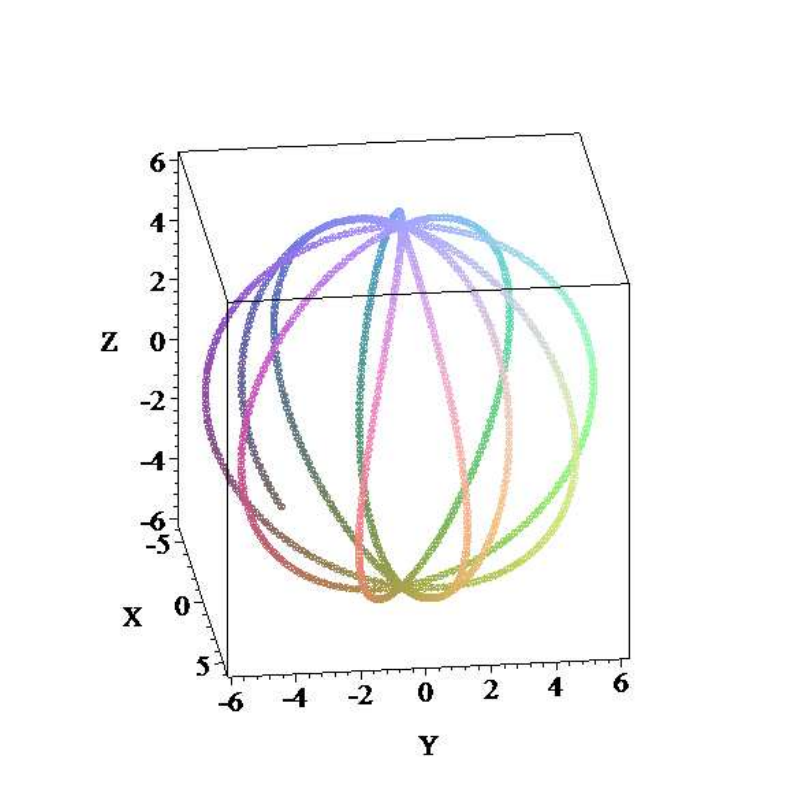}
\caption{The polar ISCO orbit in the case of $a=0.9M$ where we set $M=1$.}
\label{fig:isco}
\end{figure}

As a result, we have the ISCO radius as the following. 
\begin{widetext}
\begin{eqnarray}\label{eq:rISCOPolar}
r_{ISCO} &=& M 
\Big\{
\left(1-{\chi}^{2}+i\chi\sqrt {2-{\chi}^{2}}\right)^{1/3}
+\left(1-{\chi}^{2}+i\chi\sqrt {2-{\chi}^{2}}\right)^{-1/3}
+1 
\nonumber \\ && 
+ \sqrt{\Big[\left(1-{\chi}^{2}+i\chi\sqrt {2-{\chi}^{2}}\right)^{1/3}
+\left(1-{\chi}^{2}+i\chi\sqrt {2-{\chi}^{2}}\right)^{-1/3}
+1\Big]^2-\chi^2} 
\Big\} \,. 
\end{eqnarray}
\end{widetext}
We show the polar ISCO orbit for the case of $a=0.9M$ in Fig.~\ref{fig:isco}. 

\bibliographystyle{apsrev}
\bibliography{../../Bibtex/references}

\end{document}